\documentclass[english]{article}
\usepackage[T1]{fontenc}
\usepackage[latin9]{inputenc}
\usepackage{mathrsfs}
\usepackage{amsmath}
\usepackage{cancel}
\usepackage{esint}

\makeatletter
\usepackage{fullpage}

\makeatother

\usepackage{babel}
\begin{document}
\title{Sources of torsion in Poincarè gauge gravity}
\author{James T. Wheeler\thanks{Utah State University Dept. of Physics; jim.wheeler@usu.edu, https://orcid.org/0000-0001-9246-0079}}
\maketitle
\begin{abstract}
We study sources for torsion in Poincarè gauge theory of any dimension,
signature, and spin. We find that symmetric kinetic terms for non-Yang-Mills
bosonic fields of arbitrary rank drive torsion. Our detailed discussion
of spin-3/2 Rarita-Schwinger fields shows that they source all independent
parts of the torsion. We develop systematic notation for spin-(2k+1)/2
fields and find the spin tensor for arbitrary k in n \ensuremath{\ge}
2k+1 dimensions. For $k>0$ there is a novel direct coupling between
torsion and spinor fields. We also cast the well-known gauge relation
between the canonical and Belinfante-Rosenfield energy tensors in
terms of different choices of independent variables. 

\pagebreak{}
\end{abstract}

\section{Introduction}

The development of Riemann-Cartan geometry using the Einstein-Hilbert
action is known as the Einstein-Cartan-Sciama-Kibble (ECSK) model
of gravity. Its long history begins with Cartan's generalization of
Riemannian geometry \cite{Cartan 1922,Cartan 1923,Cartan 1924,Cartan 1925}.
A few years later Einstein used torsionful geometry to discuss teleparallel
model \cite{Einstein 1928} though this theory is not cast in the
same terms as general relativity. Originally, the evolving ECSK theory
was the study of the metric variation of the Einstein-Hilbert action
$S_{EH}\left[g\right]$ in a Riemann-Cartan geometry. The gauge theory
approach was more fully developed starting with Utiyama and continuing
with the work of Sciama and Kibble, taking its present form with the
work of Ne'eman and Regge \cite{Utiyama,Sciama,Kibble 1961,Ne'emanRegge,Ne'emanRegge2,IvanovNiederle,IvanovNiederleII}.
A detailed review is given in \cite{HehlNester}. With the advent
of modern gauge theory it has become natural to vary both metric and
connection $S_{EH}\left[g,\Gamma\right]$ or both solder form and
spin connection $S_{EH}\left[e,\omega\right]$.

Basing gravity theory on the Einstein-Hilbert action with source fields,
torsion is found to be non-propagating and vanishing away from material
sources. This is perhaps a benefit, since there is no direct experimental
evidence of torsion, and limits on torsion coupling to matter are
strong (see Donald E. Neville\footnote{``The torsion must couple to spins with coupling constants much smaller
than the electromagnetic fine-structure constant, or the force between
two macroscopic ferromagnets, due to torsion exchange, would be huge,
far greater than the familiar magnetic force due to photon exchange.''} \cite{Neville 1980}). For this reason, much study of ECSK theory
has focussed on showing that torsion does not persist in physical
situations (e.g., \cite{CarrollField}). It is natural that the seemingly
pathological non-integrability, the anomolous effect on angular momentum,
and in general the extreme success of general relativity should have
this effect. Nonetheless, the study of ECSK theory has drawn considerable
attention over the last century, including generalizations to propagating
torsion \cite{Neville 1980,SezginvanNieuwenhuizen,CarrollField,Saa,BelyaevShapiro}.
The latter have been criticized as incapable of simultaneous unitarity
and normalizability \cite{Shapiro}.

On the other hand, a deeper understanding of geometry and general
relativity is to be gained by fully exploring nearby theories. This
is the goal of the present work: to describe broad classes of sources
for torsion in Poincarè gauge theory. Our results hold in any dimension
$n$ and any signature $\left(p,q\right)$. The exercise includes
some important physical predictions, since some of the sources we
discuss, notably the spin-$\frac{3}{2}$ Rarita-Schwinger field, are
predicted by string and other supergravity theories.

In the next Section we present the basic properties of Poincarè gauge
theory using Cartan methods. We include the structure equations, Bianchi
identities, the solution for the spin connection in terms of the compatible
connection and the contorsion, and the decomposition of the torsion
into invariant parts. These results are geometrical.

The ECSK action is introduced in Section (\ref{sec:ECSK-theory}).
The variation of the solder form makes the canonical energy tensor
asymmetric. It is well-known that the choice of the local Lorentz
gauge leads to the symmetric Belinfante-Rosenfield energy tensor.
Alternatively, in the gravity case the same result is achieved using
a diffeomorphism. Here we delve a little more deeply into this equivalence,
discussing two gauge-equivalent but distinct methods of variation.
We show that the difference between the Belinfante-Rosenfield and
canonical energy tensors corresponds to a different choice of independent
variables. For the first variation the action is taken as a functional
of the solder form and the full spin connection, $S\left[\mathbf{e}^{a},\boldsymbol{\omega}_{\;\;\;b}^{a}\right]$,
in the spirit of Palatini but allowing torsion. The second variation
uses the decomposition of the spin connection into a compatible piece
and the contorsion tensor $\boldsymbol{\omega}_{\;\;\;b}^{a}=\boldsymbol{\alpha}_{\;\;\;b}^{a}+\mathbf{C}_{\;\;\;b}^{a}$.
This directly respects the Lorentz fiber structure of the bundle by
varying only the Lorentz tensors--the solder form and the contorsion,
while treating the compatible part of the spin connection as a functional
of the solder form $\boldsymbol{\alpha}_{\;\;\;b}^{a}=\boldsymbol{\alpha}_{\;\;\;b}^{a}\left(\mathbf{e}^{a}\right)$.
With matter, $\delta$$S\left[\mathbf{e}^{a},\boldsymbol{\omega}_{\;\;\;b}^{a}\right]$
leads to the canonical energy while $S\left[\mathbf{e}^{a},\mathbf{C}_{\;\;\;b}^{a}\right]$
gives the symmetric Belinfante-Rosenfeld energy tensor. The fact that
the two sets of independent variables are related by a local Lorentz
tranformation recovers the known result. The field equations in the
two cases are identical, modulo the constraints of Lorentz invariance.

The bulk of our investigation, presented in Section (\ref{sec:Sources-for-torsion}),
concerns the effects of various types of fundamental fields on torsion.
The exceptional cases of Klein-Gordon and Yang-Mills fields are treated
first. The actions for these fields do not depend on the spin connection
and therefore do not provide sources for torsion. Next, we study a
class of bosonic fields of arbitrary spin with actions quadratic and
symmetric in covariant derivatives. Except for scalars, these drive
torsion. In Subsection (\ref{subsec:Dirac-fields}) we derive the
well-known axial current source for totally antisymmetric torsion
arising from Dirac fields. We also check the effect of nonvanishing
spin tensor in the limit of general relativity where the torsion vanishes.

The effect of the less thoroughly studied Rarita-Schwinger field on
torsion is examined in Subsection (\ref{subsec:Rarita-Schwinger}).
While the axial source for Dirac fields arises from the anticommutator
of a $\gamma$-matrix with the spin connection, the Rarita-Schwinger
field couples through a similar anticommutator but with the product
of three $\gamma$-matrices. In addition, we find a new direct coupling
of the spin-$\frac{3}{2}$ field to torsion. Unlike the Dirac field
with only an axial current source, the Rarita-Schwinger field drives
all three independent pieces of the torsion. Except in dimensions
$5,7$ and $9$, spin-$\frac{3}{2}$ fields have enough degrees of
freedom to drive all components of the torsion independently.

Finally, we introduce new compact notation for spin-$\frac{2k+1}{2}$
spinor-valued $p$-form fields in Subsection (\ref{sec:Sources-for-torsion}).
This enables us to write actions for arbitrary $k$ and find the general
form of the spin tensor. The physical properties appear to echo those
of the Rarita-Schwinger field.

We conclude with a brief summary of our results.

\section{Poincare gauge theory\label{sec:Poincare-gauge-theory}}

All results below hold in arbitrary dimension $n=p+q$ and signature
$s=p-q$. The group we gauge is then $SO\left(p,q\right)$ or $Spin\left(p,q\right)$
with the familiar spacetime case having $p=3,q=1$.

\subsection{The structure of Riemann-Cartan geometry}

We review the formal features of Poincarè gauge theory. All results
below hold in arbitrary dimension $n=p+q$ and signature $s=p-q$
so while we continue to refer to the Poincarè group $ISO\left(3,1\right)$
and its Lorentz subgroup $SO\left(3,1\right)$ we actually work with
$\mathcal{P}=ISO\left(p,q\right)$ or $\mathcal{P}=ISpin\left(p,q\right)$
with subgroups $\mathcal{L}=SO\left(p,q\right)$ or $\mathcal{L}=Spin\left(p,q\right)$
respectively. The local Lorentz arena for general relativity in $n$
dimensions follows by setting $q=1$. 

The most relevant results of the Cartan construction are the 2-form
expressions for the Riemann-Cartan curvature $\boldsymbol{\mathcal{R}}_{\;\;\;b}^{a}$
and torsion $\mathbf{T}^{a}$ in terms of the solder form $\mathbf{e}^{a}$
and spin connection $\boldsymbol{\omega}_{\;\;\;b}^{a}$. 

\begin{eqnarray}
\mathbf{d}\boldsymbol{\omega}_{\;\;\;b}^{a} & = & \boldsymbol{\omega}_{\;\;\;b}^{c}\wedge\boldsymbol{\omega}_{\;\;\;c}^{a}+\boldsymbol{\mathcal{R}}_{\;\;\;b}^{a}\label{Curvature}\\
\mathbf{d}\mathbf{e}^{a} & = & \mathbf{e}^{b}\wedge\boldsymbol{\omega}_{\;\;\;b}^{a}+\mathbf{T}^{a}\label{Torsion}
\end{eqnarray}
Each of these may be expanded in the orthonormal basis
\begin{eqnarray}
\boldsymbol{\mathcal{R}}_{\;\;\;b}^{a} & = & \frac{1}{2}\mathcal{R}_{\;\;\;bcd}^{a}\mathbf{e}^{c}\land\mathbf{e}^{d}\label{Horizontal curvature}\\
\mathbf{T}^{a} & = & \frac{1}{2}T_{\;\;\;bc}^{a}\mathbf{e}^{b}\land\mathbf{e}^{c}\label{Horizontal torsion}
\end{eqnarray}
In a coordinate basis $\mathbf{T}^{a}$ is given by any antisymmetric
part of the connection.

The Bianchi identities generalize to
\begin{eqnarray}
\boldsymbol{\mathcal{D}}\mathbf{T}^{a} & = & \mathbf{e}^{b}\land\boldsymbol{\mathcal{R}}_{\;\;\;b}^{a}\label{1st RC Bianchi}\\
\boldsymbol{\mathcal{D}}\boldsymbol{\mathcal{R}}_{\;\;\;b}^{a} & = & 0\label{2nd Riemann Cartan Bianchi}
\end{eqnarray}
where the covariant exterior derivatives are given by
\begin{eqnarray*}
\boldsymbol{\mathcal{D}}\boldsymbol{\mathcal{R}}_{\;\;\;b}^{a} & = & \mathbf{d}\boldsymbol{\mathcal{R}}_{\;\;\;b}^{a}+\boldsymbol{\mathcal{R}}_{\;\;\;b}^{c}\wedge\boldsymbol{\omega}_{\;\;\;c}^{a}-\boldsymbol{\omega}_{\;\;\;b}^{c}\wedge\boldsymbol{\mathcal{R}}_{\;\;\;c}^{a}\\
\boldsymbol{\mathcal{D}}\mathbf{T}^{a} & = & \mathbf{d}\mathbf{T}^{a}+\mathbf{T}^{b}\wedge\boldsymbol{\omega}_{\;\;\;b}^{a}
\end{eqnarray*}
The frame field $\mathbf{e}^{a}$ is $\left(p,q\right)$-orthonormal,
$\left\langle \mathbf{e}^{a},\mathbf{e}^{b}\right\rangle =\eta^{ab}=diag\left(1,\ldots,1,-1,\ldots,-1\right)$,
with the connection assumed to be metric compatible
\begin{eqnarray}
\mathbf{d}\eta_{ab}-\eta_{cb}\boldsymbol{\omega}_{\;\;\;a}^{c}-\eta_{ac}\boldsymbol{\omega}_{\;\;\;b}^{c} & = & 0\label{Metric compatibility}
\end{eqnarray}
Since $\mathbf{d}\eta_{ab}=0$, the spin connection is antisymmetric,
$\boldsymbol{\omega}_{ab}=-\boldsymbol{\omega}_{ba}$.

When the connection is assumed to be compatible with the metric, Eqs.(\ref{Curvature})-(\ref{2nd Riemann Cartan Bianchi})
describe Riemann-Cartan geometry in the Cartan formalism. Note that
the Cartan-Riemann curvature, $\boldsymbol{\mathcal{R}}_{\;\;\;b}^{a}$,
differs from the Riemann curvature $\mathbf{R}_{\;\;\;b}^{a}$ by
terms dependent on the torsion. When the torsion vanishes, $\mathbf{T}^{a}=0$,
the Riemann-Cartan curvature $\boldsymbol{\mathcal{R}}_{\;\;\;b}^{a}$
reduces to the Riemann curvature $\mathbf{R}_{\;\;\;b}^{a}$ and Eqs.(\ref{Curvature})
and (\ref{Torsion}) exactly reproduce the expressions for the connection
and curvature of a general Riemannian geometry. At the same time,
Eqs.(\ref{1st RC Bianchi}) and (\ref{2nd Riemann Cartan Bianchi})
reduce to the usual first and second Bianchi identities.

The structure equations, Eqs.(\ref{Curvature}) and (\ref{Torsion}),
allow us to derive explicit forms for the connection and curvature.
From Eq.(\ref{Torsion}), the result for the spin connection is
\begin{eqnarray}
\boldsymbol{\omega}_{\;\;\;b}^{a} & = & \boldsymbol{\alpha}_{\;\;\;b}^{a}+\mathbf{C}_{\;\;\;b}^{a}\label{Connection with torsion}
\end{eqnarray}
where $\mathbf{C}_{\;\;\;b}^{a}$ is the contorsion,

\begin{equation}
\mathbf{C}_{\;\;\;b}^{a}=\frac{1}{2}\left(T_{c\;\quad b}^{\;\;\;a}+T_{\;\;\;cb}^{a}-T_{bc}^{\;\quad a}\right)\mathbf{e}^{c}\label{Contorsion}
\end{equation}
The decomposition of the connection is unique. Local Lorentz transformations
transform $\boldsymbol{\alpha}_{\;\;\;b}^{a}$ inhomogeneously in
the familiar way while torsion and contorsion are tensors. The form
of contorsion (\ref{Contorsion}) in terms of torsion is unique and
invertible.

We may recover the torsion by wedging and contracting with $\mathbf{e}^{b}$.
\begin{eqnarray*}
\mathbf{C}_{\;\;\;b}^{a}\wedge\mathbf{e}^{b} & = & \mathbf{T}^{a}
\end{eqnarray*}

The torsion now enters the curvature through the connection. Expanding
the Cartan-Riemann curvature of Eqs.(\ref{Curvature}) using Eq.(\ref{Connection with torsion})
and identifying the $\boldsymbol{\alpha}$-covariant derivative, $\mathbf{D}\mathbf{C}_{\;\;\;b}^{a}=\mathbf{d}\mathbf{C}_{\;\;\;b}^{a}-\mathbf{C}_{\;\;\;b}^{c}\land\boldsymbol{\alpha}_{\;\;\;c}^{a}-\boldsymbol{\alpha}_{\;\;\;b}^{c}\land\mathbf{C}_{\;\;\;c}^{a}$
leads to
\begin{eqnarray}
\boldsymbol{\mathcal{R}}_{\;\;\;b}^{a} & = & \mathbf{R}_{\;\;\;b}^{a}+\mathbf{D}\mathbf{C}_{\;\;\;b}^{a}-\mathbf{C}_{\;\;\;b}^{c}\land\mathbf{C}_{\;\;\;c}^{a}\label{ECSK curvature}
\end{eqnarray}
This is the Riemann-Cartan curvature expressed in terms of the Riemann
curvature and the contorsion. Note that the $\boldsymbol{\alpha}$-covariant
derivative is compatible with the solder form, $\mathbf{D}\mathbf{e}^{a}=\mathbf{d}\mathbf{e}^{a}-\mathbf{e}^{b}\land\boldsymbol{\alpha}_{\;\;\;b}^{a}=0$.
If we contract with $\mathbf{e}^{b}$ we recover the Bianchi identity.
This happens because our solution for the connection automatically
satisfies the integrability condition for the connection.

Given Eq.(\ref{ECSK curvature}) for the Cartan-Riemann curvature
in terms of the Riemannian curvature and connection, we may also expand
the generalized Bianchi identities of Eqs.(\ref{1st RC Bianchi})
and (\ref{2nd Riemann Cartan Bianchi}). The first Bianchi becomes
\begin{eqnarray*}
\mathbf{d}\mathbf{T}^{a}+\mathbf{T}^{b}\wedge\left(\boldsymbol{\alpha}_{\;\;\;b}^{a}+\mathbf{C}_{\;\;\;b}^{a}\right) & = & \mathbf{e}^{b}\wedge\mathbf{R}_{\;\;\;b}^{a}+\mathbf{e}^{b}\wedge\mathbf{D}\mathbf{C}_{\;\;\;b}^{a}-\mathbf{e}^{b}\wedge\mathbf{C}_{\;\;\;b}^{c}\land\mathbf{C}_{\;\;\;c}^{a}
\end{eqnarray*}
Using $\mathbf{D}\mathbf{e}^{a}=0$ and replacing $\mathbf{C}_{\;\;\;b}^{c}\wedge\mathbf{e}^{b}=\mathbf{T}^{c}$
leads to the Riemannian Bianchi $\mathbf{e}^{b}\wedge\mathbf{R}_{\;\;\;b}^{a}=0$.

Similarly, expanding the derivative in the second Bianchi gives
\[
0=\mathbf{D}\boldsymbol{\mathcal{R}}_{\;\;\;b}^{a}+\boldsymbol{\mathcal{R}}_{\;\;\;b}^{c}\wedge\mathbf{C}_{\;\;\;c}^{a}-\mathbf{C}_{\;\;\;b}^{c}\wedge\boldsymbol{\mathcal{R}}_{\;\;\;c}^{a}
\]
Replacing $\boldsymbol{\mathcal{R}}_{\;\;\;b}^{a}=\mathbf{R}_{\;\;\;b}^{a}+\mathbf{D}\mathbf{C}_{\;\;\;b}^{a}-\mathbf{C}_{\;\;\;b}^{c}\land\mathbf{C}_{\;\;\;c}^{a}$
throughout then using $\mathbf{e}^{b}\wedge\mathbf{C}_{\;\;\;b}^{c}=\mathbf{T}^{c}$
and $\mathbf{D}^{2}\mathbf{C}_{\;\;\;b}^{a}=\mathbf{C}_{\;\;\;b}^{c}\land\mathbf{R}_{\;\;\;c}^{a}-\mathbf{C}_{\;\;\;c}^{a}\land\mathbf{R}_{\;\;\;b}^{c}$
leads to several cancellations and finally
\begin{eqnarray*}
\mathbf{D}\mathbf{R}_{\;\;\;b}^{a} & = & 0
\end{eqnarray*}
so that the Cartan-Riemann Bianchi identities hold if and only if
the Riemann Bianchi identities hold.

The first Bianchi identity relates the triply antisymmetric part of
the curvature tensor $\boldsymbol{\mathcal{R}}_{\;\;\;b}^{a}$ to
the exterior derivative of the torsion. Expanding both sides of Eq.(\ref{1st RC Bianchi}),
antisymmetrizing and contracting leads shows that the antisymmetric
part of the Ricci-Cartan tensor is simply minus the divergence
\begin{eqnarray}
\mathcal{R}_{ab}-\mathcal{R}_{ba} & = & -\mathcal{D}_{c}\mathscr{T}_{\;\;\;ab}^{c}\label{Antisymmetric Ricci}
\end{eqnarray}
where we define
\[
\mathscr{T}_{\;\;\;bc}^{a}=T_{\;\;\;bc}^{a}-\delta_{b}^{a}T_{\;\;\;ec}^{e}+\delta_{c}^{a}T_{\;\;\;eb}^{e}
\]
For all $n>2$ this is invertible, $T_{\;\;\;bc}^{a}=\mathscr{T}_{\;\;\;bc}^{a}+\frac{1}{n-2}\left(\delta_{c}^{a}\mathscr{T}_{\;\;\;eb}^{e}-\delta_{b}^{a}\mathscr{T}_{\;\;\;ec}^{e}\right)$.
In 2-dimensions $\mathscr{T}_{\;\;\;bc}^{a}$ includes only one of
the two degrees of freedom of the torsion. Although the Ricci tensor
of the Cartan-Riemann curvature acquires an antisymmetric part there
remains only a single independent contraction. Because the curvature
is a 2-form, and the spin connection is antisymmetric, the curvature
still satisfies $\mathcal{R}_{abcd}=\mathcal{R}_{ab\left[cd\right]}=\mathcal{R}_{\left[ab\right]cd}$.

These results are geometric; a physical model follows when we posit
an action functional. The action may depend on the bundle tensors
$\mathbf{e}^{b},\mathbf{T}^{a},\boldsymbol{\mathcal{R}}_{\;\;\;b}^{a}$
and the invariant tensors $\eta_{ab}$ and $e_{ab\ldots d}$. To this
we may add source functionals built from any field representations
of the fiber symmetry group $\mathcal{L}$, including scalars, spinors,
vector fields, etc.

Constraining the torsion to zero, specifying the Einstein-Hilbert
form of action, and varying only the solder form, the $q=1$ theory
describes general relativity as a gauge theory in $n$-dimensions.
We cannot vary the metric and connection independently because this
can introduce nonzero sources for torsion, making the $\mathbf{T}^{a}=0$
constraint inconsistent.

Dropping the torsion constraint while retaining the Einstein-Hilbert
action gives the Einstein-Cartan-Sciama-Kibble (ECSK) theory of gravity
in Riemann-Cartan geometry. The torsion is found to depend on the
spin tensor, given by the connection variation of the source $\sigma$$_{\;\;\;ab}^{\mu}$$=\frac{\delta L}{\delta\omega_{\;\;\mu}^{ab}}$.
Without modifying the action to include dynamical torsion, the resulting
torsion survives only within matter.

\subsection{Decompostion of the torsion}

We identify well-known invariant pieces of the torsion. The torsion
includes a totally antisymmetric piece
\begin{eqnarray}
\mathbf{T} & \equiv & \frac{1}{3}\mathbf{e}^{a}\land\mathbf{T}_{a}=\frac{1}{3!}T_{abc}\mathbf{e}^{a}\land\mathbf{e}^{b}\land\mathbf{e}^{c}\label{Antisymmetric piece}
\end{eqnarray}
with $\frac{1}{6}n\left(n-1\right)\left(n-2\right)$ degrees of freedom.
Note that in 3, 4 or 5 dimensions the dual of $\mathbf{T}$ is a lower
rank object.
\begin{eqnarray*}
\,^{*}\mathbf{T} & = & \frac{1}{3!}T^{abc}e_{abc}\\
\,^{*}\mathbf{T} & = & \frac{1}{3!}T^{abc}e_{abcd}\mathbf{e}^{d}\\
\,^{*}\mathbf{T} & = & \frac{1}{3!2!}T^{abc}e_{abcde}\mathbf{e}^{d}\land\mathbf{e}^{e}
\end{eqnarray*}
in particular giving the well-known axial vector in 4-dimensions.
There is also a single vectorial contraction, $T_{\;\;\;ba}^{b}$.

In components the decomposition is simply
\begin{eqnarray}
T_{\;\;\;bc}^{a} & = & \tau_{\;\;\;bc}^{a}+\frac{1}{n-1}\left(\delta_{b}^{a}T_{\;\;\;ec}^{e}-\delta_{c}^{a}T_{\;\;\;eb}^{e}\right)+\eta^{ae}T_{\left[ebc\right]}\label{Decomposition in components}
\end{eqnarray}
While the vector and pseudovector each have 4 degrees of freedom in
4-dimensions, the situation is very different in higher dimensions.
In general the torsion has a total of $\frac{n^{2}\left(n-1\right)}{2}$
degrees of freedom. Therefore, while the trace contains only $n$
degrees of freedom for a fraction $\frac{2}{n\left(n-1\right)}\sim\frac{1}{n^{2}}$
of the total, the antisymmetric part includes $\frac{1}{3!}n\left(n-1\right)\left(n-2\right)$
or roughly $\frac{n-2}{3n}\sim\frac{1}{3}$. The residual tensor $\boldsymbol{\tau}^{a}$
includes the remaining $\frac{2\left(n^{2}-4\right)}{3n\left(n-1\right)}\sim\frac{2}{3}$.
Thus, the antisymmetric part is a major contributor in higher dimensions. 

\section{ECSK theory\label{sec:ECSK-theory}}

The physical content of the Einstein-Cartan-Sciama-Kibble theory enters
through use of the Einstein-Hilbert action in Riemann-Cartan geometry.

Define a volume form as the Hodge dual of unity, $\boldsymbol{\Phi}={}^{*}1=\frac{1}{n!}e_{ab\ldots c}\mathbf{e}^{a}\wedge\mathbf{e}^{b}\wedge\ldots\wedge\mathbf{e}^{c}$.
It follows that $^{*}\boldsymbol{\Phi}=\left(-1\right)^{q}$ in signature
$\left(p,q\right)$ and
\begin{eqnarray*}
\underbrace{\mathbf{e}^{a}\wedge\mathbf{e}^{b}\wedge\ldots\wedge\mathbf{e}^{c}}_{n\:terms} & = & \left(-1\right)^{q}e^{ab\ldots c}\boldsymbol{\Phi}
\end{eqnarray*}
where $e_{ab\ldots c}$ is the Levi-Civita tensor. Let $\varepsilon_{ab\ldots c}$
be the totally antisymmetric symbol with $\varepsilon_{12\ldots n}=1$
and $e=\det\left(e_{\mu}^{\;\;\;a}\right)=\sqrt{\left|g\right|}$,
so that $e_{12\ldots n}=e\varepsilon_{12\ldots n}$ and $e^{12\ldots n}=\left(-1\right)^{q}\frac{1}{e}\varepsilon_{abcd}$.

Now the ECSK action in $n$-dimensions is
\begin{equation}
S_{ECSK}\left[\mathbf{e}^{a},\boldsymbol{\omega}_{\;\;\;b}^{a}\right]=S_{EH}+S_{matter}+S_{GHY}\label{ECSK action}
\end{equation}
where
\[
S_{EH}=\frac{\kappa}{2\left(n-2\right)!}\int\boldsymbol{\mathcal{R}}^{ab}\wedge\mathbf{e}^{c}\wedge\ldots\wedge\mathbf{e}^{d}e_{abc\ldots d}=\frac{1}{2}\kappa\int\mathcal{R}\boldsymbol{\Phi}
\]
with $\boldsymbol{\mathcal{R}}^{ab}$ the Riemann-Cartan curvature
scalar, $S_{matter}$ has the general form
\[
S_{matter}=\int\mathcal{L}\left(\xi^{A},\mathcal{D}_{\mu}\xi^{A},\mathbf{e}^{a}\right)\boldsymbol{\Phi}
\]
for fields $\xi^{A}$ of any type, and $S_{GHY}$ is the Gibbons-Hawking-York
surface term
\begin{eqnarray*}
\\
\\
S_{GHY} & = & \frac{1}{8\pi}\intop_{\delta\mathcal{M}}d^{3}x\,\epsilon\,\sqrt{\left|h\right|}K
\end{eqnarray*}
where $K$ is the trace of the second fundamental form, $\epsilon=\pm1$
and $h$ the induced metric on the boundary $\delta\mathcal{M}$.

For the gravity action, we restrict attention to the Einstein-Hilbert
form but with the Riemann-Cartan scalar curvature. Alternatives with
propagating torsion are considered in \cite{Neville 1980,SezginvanNieuwenhuizen,CarrollField,Saa},
and with additional modification in \cite{Wheeler2023}.

The Gibbons-Hawking-York surface term \cite{York1972,GibbonsHawking,HawkingHorowitz,BrownYork}
is necessary because fixing both $\delta\mathbf{e}^{a}=0$ and $\delta\boldsymbol{\omega}_{\;\;\;b}^{a}=0$
on the boundary overdetermines the solution in the bulk. This can
be seen from the conditions for the initial value problem--specifying
the metric and the intrinsic curvature of an initial Cauchy surface
is enough to propagate a unique solution as the time evolution. It
is straightforward to check that adding the Gibbons-Hawking-York surface
term resolves the issue, while leaving the expected field equations
in the bulk. Having checked this, our considerations below focus on
the gravity and matter terms of the action.

The form of the field equations (and the physical content for some
nonstandard sources) also depends on the choice of independent variables.
While only the metric was varied in the original formulation in Riemann-Cartan
geometry, the gauge theory approach leads naturally to a Palatini
variation, $S\left[\mathbf{e}^{a},\boldsymbol{\omega}_{\;\;\;b}^{a}\right]$.
With this change and certain sources, it becomes impossible to set
the torsion to zero since varying the spin connection in some matter
actions gives nonvanishing sources for torsion. We explore the nature
of these sources for a variety of types of field in Section \ref{sec:Sources-for-torsion},
which contains our principal results.

Before studying sources for torsion, we examine the dependence of
the field equations on the choice of independent variables. It is
immediate that the asymmetry of the solder form means that the Einstein
tensor and canonical energy tensor may acquire antisymmetric parts
\cite{HayashiNakano}. Belinfante \cite{Belinfante} and Rosenfeld
\cite{Rosenfeld} showed that a symmetric energy tensor may be formed
by adding a combination of spin currents to the canonical energy.
The resulting Belinfante-Rosenfeld energy tensor remains divergence-free.
It is well-known that the difference between the canonical and symmetric
energy tensors can be traced to a change of local Lorentz gauge, or
in Riemannian geometry to a diffeomorphism.

Perhaps not so widely recognized is the relationship between the two
forms of energy tensor and the choice of independent variables. In
the remainder of this Section, we show first that the Palatini variation
$\delta S\left[\mathbf{e}^{a},\boldsymbol{\omega}_{\;\;\;b}^{a}\right]=0$
leads to the asymmetric form of the Ricci tensor and the canonical
energy tensor. Next, we show that the antisymmetric part of the Einstein
equation is identically satisfied as a consequence of Lorentz invariance.
Finally, we show that the alternative variation $\delta S\left[\mathbf{e}^{a},\mathbf{C}_{\;\;\;b}^{a}\right]$--allowed
by the decomposition of the connection in Eq.(\ref{Connection with torsion})--leads
to a symmetric combination of the Ricci and spin tensors for the geometry
and to the symmetric Belinfante-Rosenfeld energy tensor for matter
sources. The symmetrizing terms arise because the compatible part
of the connection must be treated as a functional of the solder form,
$\boldsymbol{\alpha}_{\;\;\;b}^{a}=\boldsymbol{\alpha}_{\;\;\;b}^{a}\left[\mathbf{e}^{a}\right]$.

\subsection{Palatini variation}

Variation of the Riemann-Cartan curvature with respect to the spin
connection leads to the $\boldsymbol{\omega}^{ab}$-covariant derivative
of the variation, $\delta\boldsymbol{\mathcal{R}}^{ab}=\boldsymbol{\mathcal{D}}\left(\delta\boldsymbol{\omega}^{ab}\right)=\mathbf{d}\left(\delta\boldsymbol{\omega}^{ab}\right)-\left(\delta\boldsymbol{\omega}^{eb}\right)\wedge\boldsymbol{\omega}_{\;\;\;e}^{a}-\left(\delta\boldsymbol{\omega}^{ae}\right)\wedge\boldsymbol{\omega}_{\;\;\;e}^{b}$.
To avoid ambiguities for surface terms, we integrate only the exterior
derivative by parts, using Lorentz invariance of the Levi-Civita tensor
to redistribute the spin connections in the remaining terms.

Varying the solder form and connection independently then leads to
\begin{eqnarray*}
\kappa\left(\mathcal{R}_{ab}-\frac{1}{2}\mathcal{R}\eta_{ab}\right) & = & \frac{\delta\mathcal{L}}{\delta e_{\mu}^{\;\;\;b}}e_{\mu}^{\;\;\;c}\eta_{ca}\\
\frac{\kappa}{2}\mathscr{T}_{\;\;\;ab}^{c} & = & -\frac{\delta\mathcal{L}}{\delta\omega_{\;\quad c}^{ab}}
\end{eqnarray*}
The asymmetric Einstein tensor is sourced by the asymmetric \emph{canonical
energy tensor} 
\begin{eqnarray}
T_{ab}^{C} & \equiv & \frac{\delta\mathcal{L}}{\delta e_{\mu}^{\;\;\;a}}\eta_{bc}e_{\mu}^{\;\;\;c}\label{Canonical energy tensor}
\end{eqnarray}
while the torsion is sourced by the \emph{spin tensor}
\begin{eqnarray}
\sigma_{\;\;\;ab}^{c} & \equiv & \frac{\delta\mathcal{L}}{\delta\omega_{\;\quad c}^{ab}}\label{Spin tensor}
\end{eqnarray}
with $\sigma_{\;\;\;ab}^{c}=-\sigma_{\;\;\;ba}^{c}$. The resulting
form of the field equations
\begin{eqnarray}
\kappa\left(\mathcal{R}_{ab}-\frac{1}{2}\mathcal{R}\eta_{ab}\right) & = & T_{\;\;\;ab}^{C}\nonumber \\
\frac{\kappa}{2}\mathscr{T}_{\;\;\;ab}^{c} & = & -\sigma_{\;\;\;ab}^{c}\label{Palatini field equations}
\end{eqnarray}
displays a pleasing completeness. The two Casimir operators of the
Poincarè group show the invariance of mass and spin, and the corresponding
energy and spin tensors provide the sources for curvature and torsion,
respectively.

In the absence of sources $\mathscr{T}_{\;\;\;ab}^{c}=0$ implies
vanishing torsion for all $n>2$, and therefore vanishing contorsion,
$\mathbf{C}_{\;\;\;c}^{a}=0$. Using Eq.(\ref{ECSK curvature}) to
separate the usual Einstein tensor from the contorsion contributions
\begin{eqnarray*}
\boldsymbol{\mathcal{R}}_{\;\;\;b}^{a} & = & \mathbf{R}_{\;\;\;b}^{a}+\mathbf{D}\mathbf{C}_{\;\;\;b}^{a}-\mathbf{C}_{\;\;\;b}^{c}\land\mathbf{C}_{\;\;\;c}^{a}
\end{eqnarray*}
and setting $\mathbf{C}_{\;\;\;c}^{a}=0$ reduces $\boldsymbol{\mathcal{R}}_{\;\;\;b}^{a}$
to the Riemann curvature $\mathbf{R}_{\;\;\;b}^{a}$ and therefore
to the usual Einstein equation of Riemannian geometry, $R_{ab}-\frac{1}{2}\eta_{ab}R=0$.
Therefore vacuum Poincarè gauge theory reproduces vacuum general relativity.
However, the theories differ when matter fields other than Yang-Mills
or Klein-Gordon type are included.

\subsection{Lorentz invariance and the symmetry of the energy tensor}

Now consider Noether's theorem applied to the action. Each term in
the total action must be Lorentz invariant, and this implies conditions
on the fields. We express these conditions for a general action.

Under local Lorentz transformation $\Lambda$, both the solder form
and spin connection change. The change in the spin connection is given
by the usual gauge form $\tilde{\boldsymbol{\omega}}=\Lambda\boldsymbol{\omega}\Lambda^{-1}-\mathbf{d}\Lambda\Lambda^{-1}$.
In detail, for an infinitesimal gauge transformation $\Lambda_{\;\;\;b}^{a}=\delta_{\;\;\;b}^{a}+\varepsilon_{\;\;\;b}^{a}$
where $\varepsilon_{ab}=-\varepsilon_{ba}$ this implies a change
in the spin connection $\delta_{\Lambda}\boldsymbol{\omega}_{\;\;\;b}^{a}=-\boldsymbol{\mathcal{D}}\varepsilon_{\;\;\;b}^{a}$
At the same time the solder form transforms as a Lorentz tensor, $\delta_{\Lambda}\mathbf{e}^{a}=\varepsilon_{\;\;\;b}^{a}\mathbf{e}^{b}$.
This means that under an infinitesimal gauge transformation we must
include changes in both the solder form and the spin connection.

Neglecting the GHY surface term the action is
\begin{eqnarray*}
S_{ECSK}\left[\mathbf{e}^{a},\boldsymbol{\omega}_{\;\;\;b}^{a}\right] & = & \frac{\kappa}{2\left(n-2\right)!}\int\boldsymbol{\mathcal{R}}^{ab}\wedge\mathbf{e}^{c}\wedge\ldots\wedge\mathbf{e}^{d}e_{abc\ldots d}\\
 &  & +\int\mathcal{L}\left(\xi^{A},\mathcal{D}_{\mu}\xi^{A},\mathbf{e}^{a}\right)\boldsymbol{\Phi}
\end{eqnarray*}
Application of the Noether theorem to each term is straightforward.
Following a general variation, we impose the field equations and restrict
the variation to the symmetry.

For the matter action we immediately have
\begin{eqnarray}
\delta S & = & \int\left(\frac{\delta\mathcal{L}}{\delta e_{\alpha}^{\;\;\;a}}\delta e_{\alpha}^{\;\;\;a}+\frac{\delta\mathcal{L}}{\delta\omega_{\;\;\quad\alpha}^{ab}}\delta\omega_{\;\;\quad\alpha}^{ab}\right)\nonumber \\
 & = & \int\left(-T_{ab}^{C}+\mathcal{D}_{c}\sigma_{\;\;\;ab}^{c}\right)\varepsilon^{ab}\label{Lorentz variation of matter action}
\end{eqnarray}
and with $\varepsilon_{ab}=-\varepsilon_{ba}$ otherwise arbitrary
this fixes the antisymmetric part of the canonical energy tensor to
equal the divergence of the spin tensor.
\begin{eqnarray}
\frac{1}{2}\left(T_{ab}^{C}-T_{ba}^{C}\right) & = & \mathcal{D}_{c}\sigma_{\;\;\;ab}^{c}\label{Lorentz invariance matter}
\end{eqnarray}

The calculation for the gravity action is somewhat more involved.
The steps result in a vanishing divergence
\begin{eqnarray*}
0 & = & \frac{\kappa}{2\left(n-2\right)!}\int\boldsymbol{\mathcal{D}}\left(-\boldsymbol{\mathcal{D}}\varepsilon^{ab}\wedge\mathbf{e}^{c}\wedge\ldots\wedge\mathbf{e}^{d}e_{abc\ldots d}\right)
\end{eqnarray*}
which requires the Ricci identity for $\boldsymbol{\mathcal{D}}^{2}\varepsilon^{ab}$
and two integrations by parts.
\begin{eqnarray*}
0 & = & -\frac{\kappa}{\left(n-2\right)!}\int\varepsilon^{ef}\delta_{e}^{a}\boldsymbol{\mathcal{R}}_{\;\;\;f}^{b}\wedge\mathbf{e}^{c}\wedge\ldots\wedge\mathbf{e}^{d}e_{abc\ldots d}\\
 &  & -\frac{\kappa}{2\left(n-3\right)!}\int\left(\varepsilon^{ab}\left(\boldsymbol{\mathcal{D}}\mathbf{T}^{c}\wedge\mathbf{e}^{d}\wedge\ldots\wedge\mathbf{e}^{e}e_{abcd\ldots e}\right)\right)\\
 &  & +\frac{\kappa}{2\left(n-4\right)!}\int\left(\varepsilon^{ab}\left(\mathbf{T}^{c}\wedge\mathbf{T}^{d}\wedge\mathbf{e}^{e}\wedge\ldots\wedge\mathbf{e}^{f}e_{abcde\ldots f}\right)\right)
\end{eqnarray*}
The final term vanishes while the first two resolve to
\begin{eqnarray*}
0 & = & -\frac{\kappa}{2}\int\varepsilon^{ab}\left(\mathcal{R}_{ab}-\mathcal{R}_{ba}\right)\boldsymbol{\Phi}\\
 &  & -\frac{\kappa}{2}\int\varepsilon^{ab}\left(\mathcal{D}_{c}T_{\;\;\;ab}^{c}+\mathcal{D}_{b}T_{\;\;\;ca}^{c}-\mathcal{D}_{a}T_{\;\;\;cb}^{c}\right)\boldsymbol{\Phi}
\end{eqnarray*}
With the arbitrariness of $\varepsilon^{ab}$ and the definition $\mathscr{T}_{\;\;\;bc}^{a}=T_{\;\;\;bc}^{a}-\delta_{b}^{a}T_{\;\;\;ec}^{e}+\delta_{c}^{a}T_{\;\;\;eb}^{e}$
this becomes
\begin{eqnarray}
\mathcal{R}_{ab}-\mathcal{R}_{ba} & = & -\mathcal{D}_{c}\mathscr{T}_{\;\;\;ab}^{c}\label{Lorentz invariance curvature}
\end{eqnarray}
which is in exact agreement with the trace of the Bianchi identity,
Eq.(\ref{Antisymmetric Ricci}).

This fully resolves the difference between asymmetric and symmetric
energy tensors. Combining the requirements for Lorentz invariance,
Eqs.(\ref{Lorentz invariance matter}) and (\ref{Lorentz invariance curvature}),
with the field equations (\ref{Palatini field equations}) shows that
the symmetric part of the Einstein tensor equals the symmetric part
of the energy source, with torsion sourced by the spin tensor as before
\begin{eqnarray}
\kappa\left(\mathcal{R}_{\left(ab\right)}-\frac{1}{2}\mathcal{R}\eta_{ab}\right) & = & T_{\left(ab\right)}^{C}\label{Curvature field equation}\\
\frac{\kappa}{2}\mathscr{T}_{\;\;\;ab}^{c} & = & -\sigma_{\;\;\;ab}^{c}\label{Torsion field equation}
\end{eqnarray}
Using the Lorentz conditions the antisymmetric part of the Einstein
equation now becomes
\begin{eqnarray*}
\frac{\kappa}{2}\mathcal{D}_{c}\mathscr{T}_{\;\;\;ab}^{c} & = & -\mathcal{D}_{c}\sigma_{\;\;\;ab}^{c}
\end{eqnarray*}
and this is now automatically satisfied as the divergence of the torsion
field equation \ref{Torsion field equation}.

\subsection{The Belinfante-Rosenfeld energy and the choice of independent variables}

In 1940 Belinfante \cite{Belinfante} and Rosenfeld \cite{Rosenfeld}
showed how to modify the canonical energy with the addition of divergences
of the spin tensor to produce a simultaneously symmetric and conserved
form of the energy. The Belinfante-Rosenfeld modification is related
to the canonical energy by Lorentz transformation. Here we derive
the Belinfante-Rosenfeld energy directly by a judicious choice of
independent variables.

Equation (\ref{Lorentz variation of matter action}) shows how Lorentz
invariance involves both the solder form and the spin connection.
This means that Lorentz transformations will mix the field equations
following from the Palatini form of variation. We can change this
using the decomposition of the spin connection into compatible $\boldsymbol{\alpha}_{\;\;\;b}^{a}$
and contorsion $\mathbf{C}_{\;\;\;b}^{a}$ pieces, Eq.(\ref{Connection with torsion}).
This combines both solder form and connection variations because compatibility
requires $\boldsymbol{\alpha}_{\;\;\;b}^{a}$ to be varied as a functional
of the solder form. This is sufficient to subsume the Lorentz conditions
(\ref{Lorentz invariance matter}) and (\ref{Lorentz invariance curvature})
derived above under the solder form variation. 

Now vary $S\left[\mathbf{e}^{a},\mathbf{C}_{\;\;\;b}^{a}\right]$.

The contorsion variation follows immediately by nothing that the $\boldsymbol{\omega}_{\;\;\;b}^{a},\boldsymbol{\alpha}_{\;\;\;b}^{a}$
and $\mathbf{C}_{\;\;\;b}^{a}$ variations are related by $\boldsymbol{\omega}_{\;\;\;b}^{a}=\boldsymbol{\alpha}_{\;\;\;b}^{a}+\mathbf{C}_{\;\;\;b}^{a}$
so that
\begin{eqnarray*}
\delta\omega_{\;\;\;bc}^{a}\mathbf{e}^{b}\wedge\mathbf{e}^{c} & = & \left.\delta\alpha_{\;\;\;bc}^{a}\mathbf{e}^{b}\wedge\mathbf{e}^{c}\right|_{C\,constant}\\
 & = & \left.\delta C_{\;\;\;bc}^{a}\mathbf{e}^{b}\wedge\mathbf{e}^{c}\right|_{\alpha\,constant}
\end{eqnarray*}
Therefore, variation of the contorsion gives the same result as the
original Palatini variation.
\begin{eqnarray}
\frac{\kappa}{2}\mathscr{T}_{\;\;\;ab}^{c} & = & -\sigma_{\;\;\;ab}^{c}\label{Torsion eq with source}
\end{eqnarray}
We may also arrive at this by expanding the curvature and varying
the $\mathbf{D}\mathbf{C}_{\;\;\;b}^{a}-\mathbf{C}_{\;\;\;b}^{c}\land\mathbf{C}_{\;\;\;c}^{a}$
terms from Eq.(\ref{ECSK curvature}) directly.

The solder form variation is now more involved. The general form is

\begin{eqnarray*}
\delta_{e}S & = & \int\left(\frac{\delta\mathcal{L}}{\delta e_{\mu}^{\;\;\;d}}+\frac{\delta\mathcal{L}}{\delta\alpha_{\;\quad c}^{ab}}\frac{\delta\alpha_{\;\quad c}^{ab}}{\delta e_{\mu}^{\;\;\;d}}\right)e_{\mu}^{\;\;\;e}\delta A_{\;\;\;e}^{d}\boldsymbol{\Phi}
\end{eqnarray*}
The first term, $\frac{\delta\mathcal{L}}{\delta e_{\mu}^{\;\;\;d}}$,
reproduces the Palatini results, i.e., the Einstein tensor $\kappa\left(\mathcal{R}_{ab}-\frac{1}{2}\mathcal{R}\eta_{ab}\right)$
and the canonical energy $T_{\;\;\;ab}^{C}$. The second term introduces
three derivatives to each side of the equation since

\begin{eqnarray}
\delta\boldsymbol{\alpha}_{\;\;\;b}^{a} & = & \frac{1}{2}\left(\delta_{d}^{a}\delta_{b}^{c}-\eta_{bd}\eta^{ac}\right)\left[D_{c}\left(\delta\mathbf{e}^{d}\right)-e_{c}^{\;\;\;\mu}\eta_{gh}\mathbf{e}^{g}D^{d}\left(\delta e_{\mu}^{\;\;\;h}\right)-e_{c}^{\;\;\;\alpha}\mathbf{D}\left(\delta e_{\alpha}^{\;\;\;d}\right)\right]\label{Connection dependence on e}
\end{eqnarray}
The Einstein-Hilbert variation leads to the replacement of the Einstein
tensor $G_{ab}=\mathcal{R}_{bc}-\frac{1}{2}\mathcal{R}\eta_{bc}$
by
\begin{eqnarray*}
\tilde{G}_{bc} & \equiv & \mathcal{R}_{bc}-\frac{1}{2}\mathcal{R}\eta_{bc}-\frac{1}{2}D^{a}\left(\mathscr{T}_{bac}+\mathscr{T}_{acb}+\mathscr{T}_{cab}\right)
\end{eqnarray*}
while the canonical energy is replaced by the Belinfante-Rosenfeld
energy tensor
\begin{eqnarray*}
T_{bc}^{BR} & = & T_{bc}^{C}+D^{a}\left(\sigma_{bac}+\sigma_{cab}-\sigma_{acb}\right)
\end{eqnarray*}

The added terms enforce the symmetry of each side. Antisymmetrizing
the gravitational terms 
\begin{eqnarray*}
\tilde{G}_{\left[bc\right]} & = & -\kappa\left(\mathcal{R}_{\left[bc\right]}-\frac{1}{2}D^{a}\mathscr{T}_{acb}\right)\equiv0
\end{eqnarray*}
by Lorentz invariance (\ref{Lorentz invariance curvature}) and for
$T_{bc}^{BR}$ we have
\begin{eqnarray*}
T_{\left[bc\right]}^{BR} & = & T_{\left[bc\right]}^{C}-D^{a}\sigma_{acb}\equiv0
\end{eqnarray*}
again directly expressing the Lorentz invariance found in Eq.(\ref{Lorentz invariance matter}).

If we write the alternative form of the curvature field equation
\begin{eqnarray}
\kappa\left(\mathcal{R}_{bc}-\frac{1}{2}\mathcal{R}\eta_{bc}\right) & = & \frac{\kappa}{2}D^{a}\left(\mathscr{T}_{bac}+\mathscr{T}_{acb}+\mathscr{T}_{cab}\right)\nonumber \\
 &  & +T_{bc}^{C}+D^{a}\left(\sigma_{bac}+\sigma_{cab}-\sigma_{acb}\right)\label{Alternative curvature field eq}
\end{eqnarray}
in symmetric and antisymmetric pieces
\begin{eqnarray*}
-\kappa\left(\mathcal{R}_{\left(bc\right)}-\frac{1}{2}\mathcal{R}\eta_{bc}-\frac{1}{2}D^{a}\left(\mathscr{T}_{bac}+\mathscr{T}_{cab}\right)\right)+T_{\left(bc\right)}^{C}+D^{a}\left(\sigma_{bac}+\sigma_{cab}\right) & = & 0\\
-\kappa\left(\mathcal{R}_{\left[bc\right]}-\frac{1}{2}D^{a}\mathscr{T}_{acb}\right)+\left(T_{\left[bc\right]}^{C}-D^{a}\sigma_{acb}\right) & = & 0
\end{eqnarray*}
we see that the added terms in the symmetrized part vanish by the
torsion field equation
\[
\frac{\kappa}{2}D^{a}\left(\mathscr{T}_{bac}+\mathscr{T}_{cab}\right)+D^{a}\left(\sigma_{bac}+\sigma_{cab}\right)=D^{a}\left(\frac{\kappa}{2}\mathscr{T}_{bac}+\sigma_{bac}\right)+D^{a}\left(\mathscr{T}_{cab}+\sigma_{cab}\right)=0
\]
while each of the two antisymmetric combinations vanishes by Lorentz
invariance, Eqs.(\ref{Lorentz invariance curvature}) and (\ref{Lorentz invariance matter})
respectively. The two variations $\delta S\left[\mathbf{e}^{a},\boldsymbol{\omega}_{\;\;\;b}^{a}\right]=0$
and $\delta S\left[\mathbf{e}^{a},\mathbf{C}_{\;\;\;b}^{a}\right]=0$
are therefore equivalent.

We therefore see that the Belinfante-Rosenfeld tensor arises directly
as the gravitational source when the independent variables are chosen
as the compatible connection and the contorsion, and the proof of
its symmetry follows from Lorentz invariance. Conservation of energy
follows as usual in general relativity, from general coordinate invariance.

\section{Sources for torsion \label{sec:Sources-for-torsion}}

We now come to our central results, the study of various sources for
torsion. Before considering fields with nonvanishing spin tensor,
we note some classes for with $\sigma_{\;\;\;bc}^{a}=0$. Fields other
than these exceptional types generically drive torsion. 

\subsection{Exceptional cases}

There are two important exceptional cases--Klein-Gordon fields and
Yang-Mills fields.

\subsubsection{Klein-Gordon field}

For Klein-Gordon fields, the covariant derivative contains no connection,
$D_{\mu}\phi=\partial_{\mu}\phi$. 
\begin{eqnarray*}
S_{KG} & = & \frac{1}{2}\int\left(g^{\mu\nu}\partial_{\mu}\phi\partial_{\nu}\phi+m^{2}\phi^{2}\right)\sqrt{\left|g\right|}d^{n}x
\end{eqnarray*}
Appropriately for a scalar field, there is no spin tensor. This holds
true for internal multiplets of scalar fields $\phi^{i}$ as well.

\subsubsection{Yang-Mills fields}

Yang-Mills fields comprise the second important class of exceptions.
Let $i,j,\ldots$ index the generators of an internal Lie symmetry
$g\in\mathcal{G}$, that is, the fiber symmetry of a principal fiber
bundle. Then the connection satisfies the Maurer-Cartan equation,
$\mathbf{d}\mathbf{A}^{i}=-\frac{1}{2}c_{\;\;jk}^{i}\mathbf{A}^{j}\land\mathbf{A}^{k}$
where $c_{\;\;jk}^{i}$ are the structure constants. Curving the bundle
the field strength
\begin{eqnarray*}
\mathbf{F}^{i} & = & \mathbf{d}\mathbf{A}^{i}+\frac{1}{2}c_{\;\;jk}^{i}\mathbf{A}^{j}\land\mathbf{A}^{k}
\end{eqnarray*}
is independent of the spacetime connection and the corresponding action
\[
S=\int\mathbf{F}^{i}\wedge\,^{*}\mathbf{F}_{i}
\]
has vanishing spin density. The result also holds for $p$-form electromagnetism
\cite{HenneauxTeitelboim} and the Proca field \cite{Proca}.

These observations mean that the Higgs and Yang-Mills fields of the
standard model do not drive torsion.

\subsection{Bosonic matter sources}

The currents of generic bosonic sources have nonvanishing spin tensors.
We consider source fields of arbitrary integer spin $\Theta^{a\ldots b}$
having quadratic kinetic energies. 

When the kinetic term of the fields is symmetric in derivatives we
have 
\[
S_{kinetic}=\frac{1}{2}\intop Q_{a\ldots bc\ldots d}\boldsymbol{\mathcal{D}}\Theta^{a\ldots b}\,^{*}\boldsymbol{\mathcal{D}}\Theta^{c\ldots d}
\]
where $Q_{a\ldots bc\ldots d}=Q_{c\ldots da\ldots b}$ for some invariant
tensor field $Q$. The contracted labels play no role in the solder
form variation, so we may write them collectively as $A=a\ldots b,B=c\ldots d$.
The action is then
\begin{eqnarray*}
S_{kinetic} & = & \frac{1}{4}\intop Q_{AB}\boldsymbol{\mathcal{D}}\Theta^{A}\,^{*}\boldsymbol{\mathcal{D}}\Theta^{B}
\end{eqnarray*}
where we assume $Q_{AB}=Q_{BA}$ is independent of the connection,
though it may depend on the metric.

The field equations (\ref{Palatini field equations}) or the reduced
equations (\ref{Curvature field equation},\ref{Torsion field equation})
hold without modification. We need only find the relevant variations
of the matter actions.

For these fields the solder form variation only enters through the
metric variation as $\eta^{ab}\left(\delta e_{a}^{\;\;\;\mu}e_{b}^{\;\;\;\nu}+e_{a}^{\;\;\;\mu}\delta e_{b}^{\;\;\;\nu}\right)=\delta g^{\mu\nu}$
since
\begin{eqnarray*}
S_{kinetic} & = & \frac{1}{2}\intop Q_{AB}\boldsymbol{\mathcal{D}}\Theta^{A}\,^{*}\boldsymbol{\mathcal{D}}\Theta^{B}\\
 & = & \frac{1}{2}\intop Q_{AB}\left(g\right)g^{\mu\nu}\mathcal{D}_{\mu}\Theta^{A}\mathcal{D}_{\nu}\Theta^{B}\sqrt{-g}d^{n}x
\end{eqnarray*}
Therefore the energy tensor takes the usual symmetric form plus any
(symmetric) dependence on $Q_{AB}$.
\begin{eqnarray*}
T_{ab} & = & Q_{AB}\mathcal{D}_{a}\Theta^{A}\mathcal{D}_{b}\Theta^{B}-\frac{1}{4}\eta_{ab}\left(Q_{AB}g^{\mu\nu}\mathcal{D}_{\mu}\Theta^{A}\mathcal{D}_{\nu}\Theta^{B}\right)+e_{a}^{\;\;\;\mu}e_{b}^{\;\;\;\nu}\frac{\delta Q_{AB}}{\delta g^{\mu\nu}}
\end{eqnarray*}
despite the asymmetric solder form variation.

However, the connection variation leads to a nonvanishing spin density.
Restoring $A\rightarrow a\ldots b,B\rightarrow c\ldots d$
\begin{eqnarray*}
\delta_{\omega}S_{kinetic} & = & \frac{1}{4\left(n-1\right)!}\delta_{\omega}\intop Q_{a\ldots bc\ldots d}\left(\mathbf{d}\Theta^{a\ldots b}+\Theta^{e\ldots b}\boldsymbol{\omega}_{\;\;\;e}^{a}+\ldots+\Theta^{a\ldots e}\boldsymbol{\omega}_{\;\;\;e}^{b}\right)\,^{*}\mathbf{D}\Theta^{c\ldots d}\\
 & = & \frac{1}{2}\intop\delta\omega_{feg}Q_{am\ldots nbc\ldots d}\left(\eta^{a[f}\Theta^{e]m\ldots nb}+\ldots+\eta^{b[f}\Theta^{\left|am\ldots n\right|e]}\right)D^{g}\Theta\:\Theta^{c\ldots d}\boldsymbol{\Phi}
\end{eqnarray*}
The spin tensor is therefore
\begin{eqnarray}
\sigma_{g}^{\;\;\;fe} & = & \frac{1}{2}Q_{am\ldots nbc\ldots d}\left(\eta^{a[f}\Phi^{e]m\ldots nb}+\ldots+\eta^{b[f}\Phi^{\left|am\ldots n\right|e]}\right)D_{g}\Phi^{c\ldots d}\label{Bosonic spin tensor}
\end{eqnarray}
This has the form of a current density.

From Lorentz invariance Eq.(\ref{Lorentz invariance matter}) and
the symmetry of the energy tensor $T_{\left[ab\right]}=0$ we immediately
have conservation of the spin tensor
\begin{eqnarray}
D_{c}\sigma_{\;\;\;ab}^{c} & = & 0\label{Conservation of spin tensor}
\end{eqnarray}
We conclude that for the types of bosonic action considered the Poincarè
gauge equations take the form
\begin{eqnarray*}
\kappa\left(\mathcal{R}_{\left(ab\right)}-\frac{1}{2}\mathcal{R}\eta_{ab}\right) & = & Q_{AB}D_{a}\Theta^{A}D_{b}\Theta^{B}-\frac{1}{2}\eta_{ab}\left(Q_{AB}g^{\mu\nu}D_{\mu}\Theta^{A}D_{\nu}\Theta^{B}\right)\\
\frac{\kappa}{2}\mathscr{T}_{c}^{\;\;\;ab} & = & -\frac{1}{2}Q_{dm\ldots nef\ldots g}\left(\eta^{d[b}\Theta^{a]m\ldots ne}+\ldots+\eta^{e[b}\Theta^{\left|dm\ldots n\right|e]}\right)D_{c}\Theta^{f\ldots g}
\end{eqnarray*}
Coupling such higher spin fields to other sources may lead to failure
of causality or other pathologies. 

For example, for a vector field with $Q_{ab}=\eta_{ab}$ the kinetic
action is simply
\[
S_{kinetic}=\frac{1}{2}\intop g^{\mu\nu}g^{\alpha\beta}\mathcal{D}_{\mu}\Theta_{\nu}\mathcal{D}_{\alpha}\Theta_{\beta}\sqrt{-g}d^{n}x
\]
so the energy tensor has the usual form and the current density is
simply $\sigma_{\mu}^{\;\;\;ab}=\frac{1}{2}\left(\Theta^{b}D_{\mu}\Theta^{a}-\Theta^{a}D_{\mu}\Theta^{b}\right)$.
The field equations are
\begin{eqnarray*}
T_{ab} & = & \eta_{cd}D_{a}\Theta^{c}D_{b}\Theta^{d}-\frac{1}{2}\eta_{ab}\left(Q_{cd}g^{\mu\nu}D_{\mu}\Theta^{c}D_{\nu}\Theta^{d}\right)\\
\sigma_{c}^{\;\;\;ab} & = & \frac{1}{4}\left(\Theta^{b}D_{c}\Theta^{a}-\Theta^{a}D_{c}\Theta^{b}\right)
\end{eqnarray*}
The torsion remains nonpropagating and vanishes whenever the source
field $\Theta^{b}$ vanishes.

\subsection{Dirac fields with torsion \label{subsec:Dirac-fields}}

It is well-known that the Dirac field provides a source for torsion
(among the earliest references see, e.g., \cite{Datta,HehlDatta,HayashiBreegman,Hehl,HehlvonderHeydeKerlick,HehlNester,HehlNitschvonderHeyde}).
The flat space Dirac action takes the same form in any dimension
\begin{eqnarray}
\mathcal{S}_{D} & = & \alpha\int\left(\bar{\psi}\left(i\cancel{\partial}-m\right)\psi\right)\,ed^{n}x\label{Riemannian Dirac action}
\end{eqnarray}
where $\cancel{\partial}=\gamma^{a}e_{a}^{\;\;\;\mu}\partial_{\mu}$.
The principal difference in dimension $n$ is that the spinors are
representations of $Spin\left(p,q\right)$ and therefore elements
of a $2^{\left[\frac{n}{2}\right]}$-dimensional complex vector space
while the $\gamma^{a}$ satisfy the Clifford algebra relations
\begin{equation}
\left\{ \gamma^{a},\gamma^{b}\right\} =-2\eta^{ab}1\label{Clifford algebra}
\end{equation}
where $\eta_{ab}$ is the $\left(p,q\right)$ metric.

However, in a curved space the spin connection introduces an additional
term. The covariant derivative of a spinor is given by
\[
D_{\mu}\psi=\partial_{\mu}\psi-\frac{1}{2}\omega_{\quad\;\mu}^{bc}\sigma_{bc}\psi
\]
where $\sigma_{bc}=\left[\gamma_{b},\gamma_{c}\right]$. The action
becomes
\begin{eqnarray*}
\tilde{\mathcal{S}}_{D} & = & \alpha\int\left(\bar{\psi}\left(i\cancel{D}-m\right)\psi\right)\,ed^{n}x\\
 & = & \alpha\int\left(\psi^{\dagger}h\left(ie_{a}^{\;\;\;\mu}\gamma^{a}D_{\mu}-m\right)\psi\right)\,ed^{n}x
\end{eqnarray*}
where $h$ is Hermitian $h^{\dagger}=h$ and reality of a vector $v^{a}=\psi^{\dagger}h\gamma^{a}\psi$
under $Spin\left(p,q\right)$ requires
\begin{eqnarray*}
\gamma^{a\dagger}h & = & h\gamma^{a}
\end{eqnarray*}
It follows that $\sigma^{ab\dagger}h=-h\sigma^{ab}$. While $h$ is
generally taken to be $\gamma^{0}$ in spacetime, $h$ transforms
as a $\left(\begin{array}{c}
0\\
2
\end{array}\right)$ spin tensor while $\gamma^{0}$ transforms as a $\left(\begin{array}{c}
1\\
1
\end{array}\right)$ spin tensor so that $h=\gamma^{0}$ can hold only in a fixed basis.
There exist satisfactory choices for $h$ in any dimension or signature
(see below). The solder form components $e_{a}^{\;\;\;\mu}$ connect
the orthonormal basis of the Clifford algebra to the coordinate basis
for the covariant derivative, $\gamma^{a}e_{a}^{\;\;\;\mu}D_{\mu}$.

The conjugate action now differs,
\begin{eqnarray*}
\tilde{\mathcal{S}}_{D}^{*} & = & \alpha\int\left(\bar{\psi}\left(-i\overleftarrow{D}_{\mu}\gamma^{\mu}-m\right)\psi\right)\,ed^{n}x
\end{eqnarray*}
so we take the manifestly real combination
\begin{eqnarray*}
\mathcal{S}_{D} & = & \frac{1}{2}\left(\tilde{\mathcal{S}}_{D}+\tilde{\mathcal{S}}_{D}^{*}\right)\\
 & = & \frac{\alpha}{2}\int\bar{\psi}\left(i\gamma^{a}\overrightarrow{\partial}_{a}-i\overleftarrow{\partial}_{a}\gamma^{a}-2m-\frac{i}{2}\omega_{bca}\left\{ \gamma^{a},\sigma^{bc}\right\} \right)\psi\,ed^{n}x
\end{eqnarray*}
showing that the connection now couples to a triple of Dirac matrices
$-\frac{i}{2}\omega_{bca}\left\{ \gamma^{a},\sigma^{bc}\right\} =-2i\omega_{bca}\gamma^{[a}\gamma^{b}\gamma^{c]}$
. This form is valid in any dimension. In 4- or 5-dimensions the triple
antisymmetrization may be shortened using $\gamma_{5}$. The action
is now
\begin{eqnarray}
\mathcal{S}_{D} & = & \alpha\int\bar{\psi}\left(\frac{i}{2}e_{a}^{\;\;\;\mu}\bar{\psi}\gamma^{a}\overleftrightarrow{\partial}_{\mu}\psi-m-ie_{a}^{\;\;\;\mu}\omega_{bc\mu}\gamma^{[a}\gamma^{b}\gamma^{c]}\right)\psi\,ed^{n}x\label{Expanded real Dirac action}
\end{eqnarray}
where $\bar{\psi}\gamma^{a}\overleftrightarrow{\partial}_{\mu}\psi=\bar{\psi}\gamma^{a}\partial_{\mu}\psi-\partial_{\mu}\bar{\psi}\gamma^{a}\psi$.

The simple form for the anticommutator turns out to be a low-dimensional
accident. In the Appendix we show that the general form for the anticommutator
$\left\{ \Gamma^{a_{1}a_{2}\ldots a_{k}},\sigma^{bc}\right\} $ depends
on both $\Gamma^{a_{1}a_{2}\ldots a_{k+1}}$ and $\Gamma^{a_{1}a_{2}\ldots a_{k-1}}$
with the second form absent for the Dirac $k=1$ case. Here we define
$\Gamma^{a_{1}a_{2}\ldots a_{k}}\equiv\gamma^{[a_{1}}\gamma^{a_{2}}\ldots\gamma^{a_{k}]}$.
This includes the particular cases $\Gamma=1$ and $\sigma^{ab}=\left[\gamma^{a},\gamma^{b}\right]$
for the $Spin\left(p,q\right)$ generators. For $k<\frac{n}{2}$ we
may write $\Gamma^{a_{1}a_{2}\ldots a_{k}}$ in terms of $\gamma_{5}\equiv i^{m}\Gamma^{a_{1}\ldots a_{n}}$
and $\Gamma^{a_{1}a_{2}\ldots a_{n-k}}$, where $i^{m}$ is chosen
so that $\gamma_{5}^{\dagger}=\gamma_{5}$.

\subsubsection{Spinor metric}

The Clifford relation for the gamma matrices is
\[
\left\{ \gamma^{a},\gamma^{b}\right\} =-2\eta^{ab}
\]
with $\eta^{ab}=diag\left(-1,\ldots,-1,1,\ldots,1\right)$. Here the
$\gamma$-matrices are numbered $\gamma^{1}\ldots\gamma^{q}\gamma^{q+1}\ldots\gamma^{q+p}$
and we take the first $q$ matrices hermitian. Then for $a,b\leq q$
the $\gamma s$ satisfy the timelike Clifford relation 
\[
\left\{ \gamma^{a},\gamma^{b}\right\} =-\eta^{ab}=+1
\]
The final $p$ $\gamma s$ must be antihermitian to give hermiticities
of $\sigma^{ab}$ appropriate for generating both rotations and boosts.

We seek a spinor metric $h$ such that both the spinor inner product
\[
\left\langle \psi,\psi\right\rangle =\psi^{\dagger A}h_{AB}\psi^{B}
\]
and the $n$-vector
\[
v^{a}\equiv\psi^{\dagger}h\gamma^{a}\psi
\]
are real. These immediately imply
\begin{eqnarray*}
h^{\dagger} & = & h\\
\gamma^{a\dagger}h & = & h\gamma^{a}
\end{eqnarray*}
To satisfy the second condition we take $h$ proportional to the product
of all timelike $\gamma s$, $h=\lambda\gamma^{1}\ldots\gamma^{q}$.
This insures that $\gamma^{a\dagger}h=\left(-1\right)^{q-1}h\gamma^{a}$
with the same sign for all $\gamma^{a}$. Then hermiticity requires
$\lambda=i^{\frac{q\left(q-1\right)}{2}}$.

This is all we need for $q$ odd. When $q$ is even we include an
additional factor of $\gamma_{5}$ where $\gamma_{5}=i^{p+\frac{n\left(n-1\right)}{2}}\gamma^{1}\ldots\gamma^{n}$.
In this case we must also include an additional $i^{q}$. Therefore
we define 
\[
h=\left\{ \begin{array}{cc}
i^{q}i^{\frac{q\left(q-1\right)}{2}}\gamma^{1}\gamma^{2}\cdots\gamma^{q}\gamma_{5} & q\;even\\
i^{\frac{q\left(q-1\right)}{2}}\gamma^{1}\gamma^{2}\cdots\gamma^{q} & q\;odd
\end{array}\right.
\]
Adopting the usual notation, we may now let $\bar{\psi}=\psi^{\dagger}h$
for spinors in any dimension. We note that $\gamma_{5}h=\left(-1\right)^{q}h\gamma_{5}$

\subsubsection{Energy tensor and spin density from the Dirac equation}

From the action (\ref{Expanded real Dirac action}) the energy tensor
and spin current are immediate. Since the Dirac Lagrangian is proportional
to the Dirac equation, there is no contribution from the volume form.
Therefore the source for the Einstein tensor is
\begin{eqnarray*}
\frac{\delta L}{\delta e_{\mu}^{\;\;\;b}}e_{\mu}^{\;\;\;c}\eta_{ca} & = & -i\alpha\bar{\psi}\gamma_{a}e_{b}^{\;\;\;\mu}\overleftrightarrow{\partial}_{\mu}\psi+2i\alpha\eta_{ac}\omega_{deb}\bar{\psi}\Gamma^{cde}\psi
\end{eqnarray*}
giving the curvature equation (\ref{Palatini field equations}) the
form
\begin{eqnarray*}
\kappa\left(R_{ab}-\frac{1}{2}R\eta_{ab}\right) & = & -i\alpha\bar{\psi}\gamma_{(a}e_{b)}^{\;\;\;\mu}\overleftrightarrow{D}_{\mu}\psi+2i\alpha\omega_{de(b}\eta_{a)c}\bar{\psi}\Gamma^{cde}\psi
\end{eqnarray*}
with $2i\alpha\omega_{de(b}\eta_{a)c}\bar{\psi}\Gamma^{cde}\psi$
becoming the axial current $\alpha\omega_{\;\quad(a}^{cd}\varepsilon_{b)cde}\bar{\psi}\gamma^{e}\gamma_{5}\psi$
in 4-dimensions.

The spin density is
\begin{eqnarray*}
\sigma^{cab} & \equiv & \frac{\delta L}{\delta\omega_{abc}}\\
 & = & -i\alpha\bar{\psi}\Gamma^{abc}\psi
\end{eqnarray*}
so the torsion is given by
\[
\frac{\kappa}{2}\mathscr{T}^{cab}=i\alpha\bar{\psi}\Gamma^{abc}\psi
\]
This is the axial current in 4-dimensions. Many studies of torsion
in ECSK and generalizations to propagating torsion are restricted
to this totally antisymmetric form of $\mathscr{T}^{cab}$.

\subsubsection{The general relativity limit\label{subsec:The-general-relativity}}

We wish to examine general relativity with coupled Dirac sources.
This source still has a spin density, despite the absence of torsion,
and it is necessary to determine whether this puts a constraint on
the Dirac field.

With vanishing torsion the connection is compatible, $\omega_{\quad\;\mu}^{bc}\rightarrow\alpha_{\quad\;\mu}^{bc}$
though the action must still be made real by adding the conjugate.
From the curvature field equation Eq.(\ref{Alternative curvature field eq})
with $\mathscr{T}_{abc}=0$, 
\begin{eqnarray*}
\kappa\left(\mathcal{R}_{bc}-\frac{1}{2}\mathcal{R}\eta_{bc}\right) & = & T_{bc}^{C}+D^{a}\left(\sigma_{bac}+\sigma_{cab}-\sigma_{acb}\right)=T_{bc}^{BR}
\end{eqnarray*}
Although there is nonvanishing spin density there is no second field
equation. There is now an antisymmetric part to the Einstein equation.
\begin{eqnarray*}
0 & = & T_{\left[bc\right]}+D^{a}\sigma_{abc}
\end{eqnarray*}
This is exactly the part that vanishes by Lorentz symmetry. The Einstein
equation therefore reduces to the symmetric expression
\begin{eqnarray*}
\kappa\left(R_{bc}-\frac{1}{2}R\eta_{bc}\right) & = & T_{\left(bc\right)}+D^{a}\left(\sigma_{bac}+\sigma_{cab}\right)
\end{eqnarray*}
where the spin tensor is the antisymmetric current
\[
\sigma^{cab}\equiv\frac{\delta L}{\delta\omega_{abc}}=-i\alpha\bar{\psi}\Gamma^{abc}\psi
\]
Because this is totally antisymmetric, $\sigma_{bac}+\sigma_{cab}=0$
and we recover the Einstein equation with the usual symmetrized energy
tensor and no additional coupling.
\begin{eqnarray*}
\kappa\left(R_{bc}-\frac{1}{2}R\eta_{bc}\right) & = & T_{\left(bc\right)}
\end{eqnarray*}
Therefore, despite nonvanishing spin tensor, Dirac fields make only
the expected contribution to the field equation of general relativity
with no additional constraint.

\subsection{Rarita-Schwinger \label{subsec:Rarita-Schwinger}}

The spin-$\frac{3}{2}$ Rarita-Schwinger field \cite{RaritaSchwinger}
is known to give rise to acausal behavior when coupled to other fields
\cite{Buchdahl}. This problem is overcome when a spin-$\frac{3}{2}$
field representing the gravitino is coupled supersymmetrically. Therefore,
we first examine the 11-dimensional supergravity Lagrangian.

\subsubsection{11-d Supergravity}

Here the basic Lagrangian
\[
\mathcal{L}=\frac{1}{2\kappa^{2}}eR-\frac{1}{2}e\overline{\psi}_{\mu}\Gamma^{\mu\nu\alpha}D_{\nu}\psi_{\alpha}+\frac{1}{48}eF_{\mu\nu\alpha\beta}^{2}
\]
includes the scalar curvature $R$, the spin-$\frac{3}{2}$ Majorana
gravitino field $\psi_{\alpha}$, and a complex 4-form field built
from a 3-form potential as $\mathbf{F}=\mathbf{d}\mathbf{A}$. The
covariant derivative has connection $\boldsymbol{\omega}_{\;\;\;b}^{a}$
and $\gamma^{\mu}=e_{a}^{\;\;\;\mu}\gamma^{a}$.

This starting Lagrangian is augmented by $\psi_{\alpha}$-$\mathbf{F}$
coupling terms and a Chern-Simons term required to enforce the supersymmetry
(\cite{Freedman vanN Rarita Sch,Freedman1976,CremmerJuliaScherk,Passiaas11D}).
The result is the Lagrangian for 11D supergravity, first found by
Cremmer, Julia and Scherk \cite{CremmerJuliaScherk}.
\begin{eqnarray*}
\mathcal{L} & = & \frac{1}{2\kappa^{2}}eR-\frac{1}{2}e\overline{\psi}_{\mu}\Gamma^{\mu\nu\alpha}D_{\nu}\left[\frac{1}{2}\left(\omega-\overline{\omega}\right)\right]\psi_{\alpha}\\
 &  & +\frac{1}{48}eF_{\mu\nu\alpha\beta}^{2}+\frac{\sqrt{2}\kappa}{384}e\left(\overline{\psi}_{\mu}\Gamma^{\mu\nu\alpha\beta\rho\sigma}\psi_{\sigma}+12\overline{\psi}^{\nu}\Gamma^{\alpha\beta}\psi^{\rho}\right)\left(F+\overline{F}\right)_{\nu\alpha\beta\rho}\\
 &  & +\frac{\sqrt{2}\kappa}{3456}\varepsilon^{\alpha_{1}\dots\alpha_{11}}F_{\alpha_{1}\dots\alpha_{4}}F_{\alpha_{5}\dots\alpha_{8}}A_{\alpha_{9}\alpha_{10}\alpha_{11}}
\end{eqnarray*}

Since we are primarily interested in sources for torsion, we will
only need the kinetic term for the Rarita-Schwinger field. While is
it possible that supergravity theories--which exist only in certain
dimensions--are the only consistent formulation of spin-$\frac{3}{2}$
fields, there may be alternative couplings that allow them. For this
reason, we will consider the original Rarita-Schwinger kinetic term
in arbitrary dimension as a source for torsion, omitting additional
couplings. 

\subsubsection{The Rarita-Schwinger equation}

In flat 4-dimensional space the uncoupled Rarita-Schwinger equation
may be written as
\begin{eqnarray*}
\varepsilon^{\mu\nu\alpha\beta}\gamma_{\nu}\gamma_{5}\partial_{\alpha}\psi_{\beta}+\frac{1}{2}m\sigma^{\mu\beta}\psi_{\beta} & = & 0
\end{eqnarray*}
with real action
\begin{eqnarray*}
S_{RS}^{0} & = & \int\bar{\psi}_{\mu}\left(\epsilon^{\mu\kappa\rho\nu}\gamma_{5}\gamma_{\kappa}\partial_{\rho}-\frac{1}{2}m\sigma^{\mu\nu}\right)\psi_{\nu}
\end{eqnarray*}

In curved spacetime, generalizing to the covariant derivative $\partial_{\alpha}\psi_{\beta}\rightarrow\mathcal{D}_{\alpha}\psi_{\beta}$
where
\begin{eqnarray*}
\mathcal{D}_{\alpha}\psi_{\beta} & = & \partial_{\alpha}\psi_{\beta}-\psi_{\mu}\Gamma_{\;\;\;\beta\alpha}^{\mu}-\frac{1}{2}\omega_{ab\alpha}\sigma^{ab}\psi_{\beta}
\end{eqnarray*}
we must explicitly make it real. As with the Dirac field, the extra
terms give an anticommutator. Noticing that
\begin{eqnarray*}
\varepsilon^{\mu\kappa\alpha\nu}\Gamma_{\;\;\;\nu\alpha}^{\rho} & = & \frac{1}{2}\varepsilon^{\mu\kappa\alpha\nu}T_{\;\;\;\alpha\nu}^{\rho}
\end{eqnarray*}
we have
\begin{eqnarray*}
S_{RS} & = & \frac{1}{2}\left(S+S^{*}\right)\\
 & = & S_{RS}^{0}-\frac{1}{2}\int\left(\epsilon^{\mu\kappa\alpha\nu}\left(\frac{1}{2}\bar{\psi}_{\mu}\gamma_{5}\gamma_{\kappa}\psi_{\rho}T_{\;\;\;\alpha\nu}^{\rho}+\frac{1}{2}\left[\bar{\psi}_{\mu}\gamma_{5}\gamma_{\kappa}\psi_{\rho}T_{\;\;\;\alpha\nu}^{\rho}\right]^{\dagger}\right)\right)\\
 &  & +\frac{1}{2}\int\left(-\frac{1}{2}\omega_{ab\alpha}\epsilon^{\mu\kappa\alpha\nu}\left(\bar{\psi}_{\mu}\gamma_{5}\gamma_{\kappa}\sigma^{ab}\psi_{\nu}+\left[\bar{\psi}_{\mu}\gamma_{5}\gamma_{\kappa}\sigma^{ab}\psi_{\nu}\right]^{\dagger}\right)\right)
\end{eqnarray*}
and therefore, taking the adjoint and rearranging
\begin{eqnarray*}
S_{RS} & = & S_{RS}^{0}-\frac{1}{4}\int\epsilon^{\mu\kappa\alpha\nu}\left(\bar{\psi}_{\mu}\gamma_{5}\gamma_{\kappa}\psi_{\rho}T_{\;\;\;\alpha\nu}^{\rho}+\bar{\psi}_{\rho}T_{\;\;\;\alpha\nu}^{\rho}\gamma_{5}\gamma_{\kappa}\psi_{\mu}\right)\\
 &  & -\frac{1}{4}\int\omega_{ab\alpha}\epsilon^{\mu\kappa\alpha\nu}\bar{\psi}_{\mu}\gamma_{5}\left\{ \gamma_{\kappa},\sigma^{ab}\right\} \psi_{\nu}
\end{eqnarray*}

The explicit torsion coupling here is surprising, and forces us to
be clear about the independent variables. We may set $\mathbf{T}^{a}=\mathbf{d}\mathbf{e}^{a}-\mathbf{e}^{b}\wedge\boldsymbol{\omega}_{\;\;\;b}^{a}$
and vary $\left(\mathbf{e}^{a},\boldsymbol{\omega}_{\;\;\;b}^{a}\right)$
or we may write $\boldsymbol{\omega}_{\;\;\;b}^{a}=\boldsymbol{\alpha}_{\;\;\;b}^{a}\left(\mathbf{e}^{c}\right)+\mathbf{C}_{\;\;\;b}^{a}$
and write the torsion in terms of the contorsion $\mathbf{T}^{a}=\mathbf{C}_{\;\;\;b}^{a}\wedge\mathbf{e}^{b}$,
then vary $\left(\mathbf{e}^{a},\mathbf{C}_{\;\;\;b}^{a}\right)$.
We choose the latter course, since this respects the Lorentz fiber
symmetry and yields the Belinfante-Rosenfield tensor as source. For
the spin tensor it makes no difference because
\begin{eqnarray*}
\delta_{\omega}\mathbf{T}^{a} & = & -\mathbf{e}^{b}\wedge\delta\boldsymbol{\omega}_{\;\;\;b}^{a}\\
\delta_{C}\mathbf{T}^{a} & = & -\mathbf{e}^{b}\wedge\delta\mathbf{C}_{\;\;\;b}^{a}
\end{eqnarray*}

Before carrying out the variation we develop Rarita-Schwinger action
in higher dimensions.

\subsubsection{The Rarita-Schwinger action in arbitrary dimension}

To explore higher dimensions we introduce some general notation. Clearly
we will need the Hodge dual, but it yields a more systematic result
if we combine the dual with the gamma matrices. 

Define:
\begin{eqnarray*}
\boldsymbol{\gamma} & \equiv & \gamma_{a}\mathbf{e}^{a}\\
\boldsymbol{\psi} & \equiv & \psi_{a}\mathbf{e}^{a}\\
\left(\land\boldsymbol{\gamma}\right)^{k} & \equiv & \gamma_{a_{1}}\ldots\gamma_{a_{k}}\mathbf{e}^{a_{1}}\wedge\ldots\wedge\mathbf{e}^{a_{k}}\\
\boldsymbol{\Gamma}^{k} & \equiv & \,^{*}\left[\frac{1}{k!}\left(\wedge\boldsymbol{\gamma}\right)^{k}\right]
\end{eqnarray*}
In particular, $\boldsymbol{\Gamma}^{0}$ is just the volume form
$\boldsymbol{\Phi}$.

It is not hard to check that the Dirac case may be written as
\begin{eqnarray*}
S_{D}^{0} & = & \intop\left(\bar{\psi}\boldsymbol{\Gamma}^{1}\land i\mathbf{d}\psi-m\bar{\psi}\boldsymbol{\Gamma}^{0}\psi-\frac{i}{4}\left\{ \boldsymbol{\Gamma}^{1},\sigma^{cd}\right\} \land\boldsymbol{\omega}_{cd}\psi\right)
\end{eqnarray*}
by expanding the forms.

To rewrite the Rarita-Schwinger action in arbitrary dimensions we
replace the volume form and set
\[
\sigma^{\mu\nu}=\left(-1\right)^{q}\left(\frac{1}{2}\sigma^{\rho\sigma}\right)\left(\frac{1}{2}e^{\alpha\beta\mu\nu}e_{\alpha\beta\rho\sigma}\right)
\]
Then
\begin{eqnarray*}
S_{RS}^{0} & = & \int\bar{\psi}_{\mu}\left(\epsilon^{\mu\kappa\rho\nu}\gamma_{5}\gamma_{\kappa}\partial_{\rho}-\frac{1}{2}m\sigma^{\mu\nu}\right)\psi_{\nu}\boldsymbol{\Phi}\\
 & = & \int\left(\epsilon^{\mu\kappa\rho\nu}\bar{\psi}_{\mu}\gamma_{5}\gamma_{\kappa}\partial_{\rho}\psi_{\nu}\frac{\left(-1\right)^{q}}{4!}e_{defg}\mathbf{e}^{d}\wedge\mathbf{e}^{e}\wedge\mathbf{e}^{f}\wedge\mathbf{e}^{g}\right)\\
 &  & -\int\frac{1}{2}m\bar{\psi}_{\mu}\left(-1\right)^{q}\frac{1}{2}\sigma^{\rho\sigma}\frac{1}{2}e^{\alpha\beta\mu\nu}e_{\alpha\beta\rho\sigma}\psi_{\nu}\frac{1}{4!}e_{defg}\mathbf{e}^{d}\wedge\mathbf{e}^{e}\wedge\mathbf{e}^{f}\wedge\mathbf{e}^{g}
\end{eqnarray*}
This allows us to eliminate the 4-dimensional Levi-Civita tensor by
reducing the Levi-Civita pairs $\frac{\left(-1\right)^{q}}{4!}\epsilon^{\mu\kappa\rho\nu}e_{defg}$
and $\frac{\left(-1\right)^{q}}{4!}e^{\alpha\beta\mu\nu}e_{defg}$,
to combine a solder form with each spinor. Then
\begin{eqnarray*}
S_{RS}^{0} & = & \int\bar{\boldsymbol{\psi}}\wedge\gamma_{5}\gamma_{e}\wedge\mathbf{e}^{e}\mathbf{d}\boldsymbol{\psi}\\
 &  & -\int\frac{1}{8}m\bar{\boldsymbol{\psi}}\wedge\left(\frac{1}{8}\sigma^{\rho\sigma}e_{\rho\sigma de}\mathbf{e}^{d}\wedge\mathbf{e}^{e}\right)\wedge\boldsymbol{\psi}
\end{eqnarray*}
Now set
\begin{eqnarray*}
\gamma_{5}\gamma_{\kappa}\mathbf{e}^{\kappa} & = & \frac{i}{3!}\gamma_{[a}\gamma_{b}\gamma_{c]}\varepsilon_{\;\;\quad\kappa}^{abc}\mathbf{e}^{\kappa}=i\boldsymbol{\Gamma}^{3}
\end{eqnarray*}
and
\begin{eqnarray*}
\frac{1}{8}\sigma^{ab}e_{abcd}\mathbf{e}^{c}\wedge\mathbf{e}^{d} & = & \frac{1}{2!}\,^{*}\left(\gamma^{a}\gamma^{b}\mathbf{e}_{a}\wedge\mathbf{e}_{b}\right)=\boldsymbol{\Gamma}^{2}
\end{eqnarray*}
to write the action as
\begin{eqnarray}
S_{RS}^{0} & = & \int\left(\bar{\boldsymbol{\psi}}\wedge\boldsymbol{\Gamma}^{3}\wedge i\mathbf{d}\boldsymbol{\psi}-m\bar{\boldsymbol{\psi}}\wedge\boldsymbol{\Gamma}^{2}\wedge\boldsymbol{\psi}\right)\label{Vacuum RS}
\end{eqnarray}
By using the Hodge dual in $\boldsymbol{\Gamma}^{2}$ and $\boldsymbol{\Gamma}^{3}$
we have eliminated the specific reference to dimension. Equation (\ref{Vacuum RS})
is the Rarita-Schwinger action in flat $\left(p,q\right)$-space.

\subsubsection{Rarita-Schwinger in curved spaces}

To generalize Eq.(\ref{Vacuum RS}) we now replace the exterior derivative
with the covariant exterior derivative
\[
\tilde{S}_{RS}=\int\left(\bar{\boldsymbol{\psi}}\wedge\boldsymbol{\Gamma}^{3}\wedge i\boldsymbol{\mathcal{D}}\boldsymbol{\psi}-m\bar{\boldsymbol{\psi}}\wedge\boldsymbol{\Gamma}^{2}\wedge\boldsymbol{\psi}\right)
\]
keeping the action real by taking $S_{RS}=\frac{1}{2}\left(\tilde{S}_{RS}+\tilde{S}_{RS}^{\dagger}\right)$.
The covariant derivative 2-form $\boldsymbol{\mathcal{D}}\boldsymbol{\psi}$
is
\begin{eqnarray}
\boldsymbol{\mathcal{D}}\boldsymbol{\psi} & = & \mathbf{d}\boldsymbol{\psi}-\psi_{\mu}\mathbf{T}^{\mu}-\frac{1}{2}\boldsymbol{\omega}_{mn}\sigma^{mn}\wedge\boldsymbol{\psi}\label{RS covariant derivative}
\end{eqnarray}
Therefore, the direct torsion-Rarita-Schwinger coupling will occur
in higher dimensions as well.

Expanding the action and separating the free contribution

\begin{eqnarray*}
S_{RS} & = & S_{RS}^{0}+\frac{1}{2}\intop\left(\bar{\boldsymbol{\psi}}\land\boldsymbol{\Gamma}^{3}\land\left(-i\psi_{\mu}\mathbf{T}^{\mu}\right)+\left(\bar{\boldsymbol{\psi}}\land\boldsymbol{\Gamma}^{3}\land\left(-i\psi_{\mu}\mathbf{T}^{\mu}\right)\right)^{\dagger}\right)\\
 &  & +\frac{1}{2}\intop\left(\bar{\boldsymbol{\psi}}\land\boldsymbol{\Gamma}^{3}\land\left(-\frac{i}{2}\boldsymbol{\omega}_{mn}\wedge\sigma^{mn}\boldsymbol{\psi}\right)+\left(\bar{\boldsymbol{\psi}}\land\boldsymbol{\Gamma}^{3}\land\left(-\frac{i}{2}\boldsymbol{\omega}_{mn}\wedge\sigma^{mn}\boldsymbol{\psi}\right)\right)^{\dagger}\right)
\end{eqnarray*}
The conjugate torsion piece us given by
\[
\frac{1}{2}\intop\left(\bar{\boldsymbol{\psi}}\land\boldsymbol{\Gamma}^{3}\land\left(-i\psi_{m}\mathbf{T}^{m}\right)\right)^{\dagger}=-\frac{i}{2}\intop\left(-1\right)^{n+1}\bar{\psi}_{m}\mathbf{T}^{m}\land\boldsymbol{\Gamma}^{3}\land\boldsymbol{\psi}
\]
and the conjugate spin connection piece becomes
\begin{eqnarray*}
\frac{1}{2}\intop\left(\bar{\boldsymbol{\psi}}\land\boldsymbol{\Gamma}^{3}\land\left(-\frac{i}{2}\boldsymbol{\omega}_{mn}\wedge\sigma^{mn}\boldsymbol{\psi}\right)\right)^{\dagger} & = & \frac{1}{2}\intop\left[\bar{\boldsymbol{\psi}}\land\sigma^{mn}\boldsymbol{\Gamma}^{3}\left(-\frac{i}{2}\land\boldsymbol{\omega}_{mn}\right)\land\boldsymbol{\psi}\right]
\end{eqnarray*}

Therefore, the full action is
\begin{eqnarray*}
S_{RS} & = & \intop\left(\bar{\boldsymbol{\psi}}\land\boldsymbol{\Gamma}^{3}\land i\mathbf{d}\boldsymbol{\psi}-m\bar{\boldsymbol{\psi}}\land\boldsymbol{\Gamma}^{2}\land\boldsymbol{\psi}\right)\\
 &  & -\frac{i}{2}\intop\left(\bar{\boldsymbol{\psi}}\land\boldsymbol{\Gamma}^{3}\land\mathbf{T}^{a}\psi_{a}-\left(-1\right)^{n}\mathbf{T}^{a}\bar{\psi}_{a}\land\boldsymbol{\Gamma}^{3}\wedge\boldsymbol{\psi}\right)\\
 &  & -\frac{i}{4}\intop\bar{\boldsymbol{\psi}}\land\left\{ \boldsymbol{\Gamma}^{3},\sigma^{cd}\right\} \land\boldsymbol{\omega}_{cd}\land\boldsymbol{\psi}
\end{eqnarray*}
The anticommutator is
\begin{eqnarray*}
\left\{ \gamma^{[a_{1}}\gamma^{a_{2}}\gamma^{a_{3}]},\sigma^{de}\right\}  & = & 4\sum_{a_{1}<a_{2}<a_{3}}\left(\gamma^{[a_{1}}\gamma^{a_{2}}\gamma^{a_{3}}\gamma^{d}\gamma^{e]}-\left(\eta^{a_{1}d}\eta^{a_{2}e}-\eta^{a_{2}d}\eta^{a_{1}e}\right)\eta^{dd}\eta^{ee}\gamma^{a_{3}}\right.\\
 &  & +\left.\left(\eta^{a_{1}d}\eta^{a_{3}e}-\eta^{a_{3}d}\eta^{a_{1}e}\right)\eta^{dd}\eta^{ee}\gamma^{a_{2}}-\left(\eta^{a_{2}d}\eta^{a_{3}e}-\eta^{a_{3}d}\eta^{a_{2}e}\right)\eta^{dd}\eta^{ee}\gamma^{a_{1}}\right)
\end{eqnarray*}
so the Rarita-Schwinger spin tensor contains couplings involving $\boldsymbol{\Gamma}^{1},\boldsymbol{\Gamma}^{3},\boldsymbol{\Gamma}^{5}$.

\subsubsection{The Rarita-Schwinger spin tensor}

Varying the action with respect to the spin connection or contorsion
\begin{eqnarray*}
\delta_{\omega}S_{RS} & = & -\intop\frac{i}{2}\bar{\boldsymbol{\psi}}\land\boldsymbol{\Gamma}^{3}\land\left(-\mathbf{e}^{b}\wedge\delta\boldsymbol{\omega}_{\;\;\;b}^{a}\right)\psi_{a}-\frac{i}{2}\intop\left(-1\right)^{n+1}\left(-\mathbf{e}^{b}\wedge\delta\boldsymbol{\omega}_{\;\;\;b}^{a}\right)\bar{\psi}_{a}\land\boldsymbol{\Gamma}^{3}\wedge\boldsymbol{\psi}\\
 &  & +\frac{i}{4}\intop\bar{\boldsymbol{\psi}}\land\left\{ \boldsymbol{\Gamma}^{3},\sigma_{\;\;\;a}^{b}\right\} \land\delta\boldsymbol{\omega}_{\;\;\;b}^{a}\land\boldsymbol{\psi}
\end{eqnarray*}
Expanding the forms, setting $\delta\boldsymbol{\omega}_{\;\;\;b}^{a}=A_{\;\;\;bc}^{a}\mathbf{e}^{c}$,
and collecting the basis into volume forms this becomes

\begin{eqnarray*}
\delta_{\omega}S_{RS} & = & \frac{i}{2}\intop A_{abc}\left(\bar{\psi}_{e}\gamma^{[e}\gamma^{b}\gamma^{c]}\psi^{a}-\bar{\psi}^{a}\gamma^{[b}\gamma^{e}\gamma^{c]}\psi_{e}-\frac{1}{2}\bar{\psi}_{d}\left\{ \gamma^{[d}\gamma^{e}\gamma^{c]},\sigma^{ba}\right\} \psi_{e}\right)\boldsymbol{\Phi}
\end{eqnarray*}
so antisymmetrizing on $ab$ and expanding the anticommutator as
\begin{eqnarray*}
\frac{i}{4}\bar{\psi}_{d}\left\{ \gamma^{[d}\gamma^{e}\gamma^{c]},\sigma^{ab}\right\} \psi_{e} & = & i\bar{\psi}_{d}\gamma^{[a}\gamma^{b}\gamma^{c}\gamma^{d}\gamma^{e]}\psi_{e}+i\left(\eta^{ac}\eta^{bd}-\eta^{bc}\eta^{ad}\right)\bar{\psi}_{d}\gamma^{e}\psi_{e}\\
 &  & +i\left(\eta^{ae}\eta^{bc}-\eta^{ac}\eta^{be}\right)\bar{\psi}_{d}\gamma^{d}\psi_{e}+i\left(\eta^{ad}\eta^{be}-\eta^{ae}\eta^{bd}\right)\bar{\psi}_{d}\gamma^{c}\psi_{e}
\end{eqnarray*}
the spin tensor is
\begin{eqnarray}
\sigma^{cab} & = & \frac{i}{4}\left(\bar{\psi}_{e}\gamma^{[e}\gamma^{b}\gamma^{c]}\psi^{a}-\bar{\psi}_{e}\gamma^{[e}\gamma^{a}\gamma^{c]}\psi^{b}+\bar{\psi}^{b}\gamma^{[a}\gamma^{e}\gamma^{c]}\psi_{e}-\bar{\psi}^{a}\gamma^{[b}\gamma^{e}\gamma^{c]}\psi_{e}\right)\nonumber \\
 &  & +i\bar{\psi}_{d}\gamma^{[a}\gamma^{b}\gamma^{c}\gamma^{d}\gamma^{e]}\psi_{e}+i\left(\eta^{ac}\eta^{bd}-\eta^{bc}\eta^{ad}\right)\bar{\psi}_{d}\gamma^{e}\psi_{e}\nonumber \\
 &  & +i\left(\eta^{ae}\eta^{bc}-\eta^{ac}\eta^{be}\right)\bar{\psi}_{d}\gamma^{d}\psi_{e}+i\left(\eta^{ad}\eta^{be}-\eta^{ae}\eta^{bd}\right)\bar{\psi}_{d}\gamma^{c}\psi_{e}\label{RS spin tensor}
\end{eqnarray}

After using the torsion equation, the source for the Einstein tensor
is always the symmetrized canonical tensor (\ref{Curvature field equation})
but the torsion is now driven by much more than the axial current.
We next use the full spin tensor , Eq.(\ref{RS spin tensor}), to
compute the source for each indepdendent part of the torsion. Since
the reduced field equation shows that $\frac{\kappa}{2}\mathscr{T}_{\;\;\;ab}^{c}=-\sigma_{\;\;\;ab}^{c}$
it suffices to find the trace, totally antisymmetric, and traceless,
mixed symmetry parts of $\sigma^{cab}$. The corresponding parts of
the torsion are proportional to these.

First, the trace of the spin tensor reduces to a simple vector current.
\begin{eqnarray*}
\sigma_{c}^{\;\;\;cb} & = & i\left(n-2\right)\left(\bar{\psi}^{b}\gamma^{e}\psi_{e}-\bar{\psi}_{e}\gamma^{e}\psi^{b}\right)
\end{eqnarray*}

For the antisymmetric part there is no change in the totally antisymmetric
piece $i\bar{\psi}_{d}\gamma^{[a}\gamma^{b}\gamma^{c}\gamma^{d}\gamma^{e]}\psi_{e}$.
Of the last three terms involving metrics, the first two vanish while
the antisymmetrization of the third gives
\begin{eqnarray*}
\left(i\left(\eta^{ad}\eta^{be}-\eta^{ae}\eta^{bd}\right)\bar{\psi}_{d}\gamma^{c}\psi_{e}\right)_{\left[abc\right]} & = & -2i\bar{\psi}^{[a}\gamma^{b}\psi^{c]}
\end{eqnarray*}
The remaining terms require the $abc$ antisymmetrization of $\bar{\psi}_{e}\gamma^{[e}\gamma^{b}\gamma^{c]}\psi^{a}$
and $\bar{\psi}^{b}\gamma^{[a}\gamma^{e}\gamma^{c]}\psi_{e}$. This
is complicated by the existing antisymmetry of $ebc$. Write these
out in detail and collecting terms we find 
\begin{eqnarray*}
\left(\bar{\psi}_{e}\gamma^{[e}\gamma^{b}\gamma^{c]}\psi^{a}\right)_{\left[abc\right]} & = & \frac{4}{3}\bar{\psi}_{e}\gamma^{[e}\gamma^{a}\gamma^{b}\psi^{c]}+\frac{1}{3}\bar{\psi}_{e}\gamma^{[a}\gamma^{b}\gamma^{c]}\psi^{e}\\
\left(\bar{\psi}^{a}\gamma^{[b}\gamma^{e}\gamma^{c]}\psi_{e}\right)_{\left[abc\right]} & = & \frac{4}{3}\bar{\psi}^{[a}\gamma^{b}\gamma^{c}\gamma^{e]}\psi_{e}+\frac{1}{3}\bar{\psi}^{e}\left(\gamma^{[a}\gamma^{b}\gamma^{c]}\right)\psi_{e}
\end{eqnarray*}
with the full contribution to $\sigma^{\left[abc\right]}$ being $\frac{i}{2}$
times these. Combining everything, the source for the totally antisymmetric
part of the torsion is

\begin{eqnarray*}
\sigma^{[cab]} & = & i\bar{\psi}_{d}\gamma^{[a}\gamma^{b}\gamma^{c}\gamma^{d}\gamma^{e]}\psi_{e}+\frac{2i}{3}\left(\bar{\psi}_{e}\gamma^{[e}\gamma^{a}\gamma^{b}\psi^{c]}-\bar{\psi}^{[a}\gamma^{b}\gamma^{c}\gamma^{e]}\psi_{e}\right)\\
 &  & +\frac{i}{6}\left(\bar{\psi}_{e}\gamma^{[a}\gamma^{b}\gamma^{c]}\psi^{e}-\bar{\psi}^{e}\left(\gamma^{[a}\gamma^{b}\gamma^{c]}\right)\psi_{e}\right)-2i\bar{\psi}^{[a}\gamma^{b}\psi^{c]}
\end{eqnarray*}
containing $1$-, $3$-, and $5$-gamma currents.

The traceless, mixed symmetry part $\tilde{\sigma}^{cab}$ is found
by subtracting the trace and antisymmetric pieces.
\[
\tilde{\sigma}^{cab}=\sigma^{cab}-\sigma^{\left[cab\right]}-\frac{1}{n-1}\left(\eta^{ac}\sigma_{c}^{\;\;\;cb}-\eta^{bc}\sigma_{c}^{\;\;\;ca}\right)
\]

The result is
\begin{eqnarray*}
\tilde{\sigma}^{cab} & = & \frac{i}{4}\left(\bar{\psi}_{e}\gamma^{[e}\gamma^{b}\gamma^{c]}\psi^{a}-\bar{\psi}_{e}\gamma^{[e}\gamma^{a}\gamma^{c]}\psi^{b}+\bar{\psi}^{b}\gamma^{[a}\gamma^{e}\gamma^{c]}\psi_{e}-\bar{\psi}^{a}\gamma^{[b}\gamma^{e}\gamma^{c]}\psi_{e}\right)\\
 &  & -\frac{2i}{3}\left(\bar{\psi}_{e}\gamma^{[e}\gamma^{a}\gamma^{b}\psi^{c]}-\bar{\psi}^{[a}\gamma^{b}\gamma^{c}\gamma^{e]}\psi_{e}\right)+2i\bar{\psi}^{[a}\gamma^{b}\psi^{c]}\\
 &  & +\frac{i}{n-1}\eta^{ac}\left(\bar{\psi}^{b}\gamma^{e}\psi_{e}-\bar{\psi}_{e}\gamma^{e}\psi^{b}\right)-\frac{i}{n-1}\eta^{bc}\left(\bar{\psi}^{a}\gamma^{e}\psi_{e}-\bar{\psi}_{e}\gamma^{e}\psi^{a}\right)+i\left(\bar{\psi}^{a}\gamma^{c}\psi^{b}-\bar{\psi}^{b}\gamma^{c}\psi^{a}\right)
\end{eqnarray*}
The traceless, mixed symmetry piece therefore depends on $1$- and
$3$-gamma currents.

Therefore, while the Dirac field produces only an axial vector source
for torsion, the Rarita-Schwinger field provides a source for each
independent piece. Moreover, since a spin-$\frac{3}{2}$ field in
$n$-dimensions has $n\times2^{\left[\frac{n}{2}\right]+1}$ degrees
of freedom while the torsion has $\frac{1}{2}n^{2}\left(n-1\right)$,
generic solutions may be expected to produce generic torsion except
in dimensions $n=5,7$ or $9$.

\subsection{Higher spin fermions}

We have seen that the vacuum Dirac $\left(k=0\right)$ and Rarita-Schwinger
$\left(k=1\right)$ actions for spin-$\frac{2k+1}{2}$ may be written
as
\begin{eqnarray*}
S_{k=0}^{0} & = & \intop\left(\bar{\psi}\boldsymbol{\Gamma}^{1}\land i\mathbf{d}\psi-m\bar{\psi}\boldsymbol{\Gamma}^{0}\psi\right)\\
S_{k=1}^{0} & = & \int\left(\bar{\boldsymbol{\psi}}\wedge\boldsymbol{\Gamma}^{3}\wedge i\mathbf{d}\boldsymbol{\psi}-m\bar{\boldsymbol{\psi}}\wedge\boldsymbol{\Gamma}^{2}\wedge\boldsymbol{\psi}\right)
\end{eqnarray*}
The pattern seen here generalizes immediately to higher fermionic
spins in any dimension $n\geq2k+1$, with the flat space kinetic term
depending on $\boldsymbol{\Gamma}^{2k+1}$ and the mass term depending
on $\boldsymbol{\Gamma}^{2k}$ for spin $\frac{2k+1}{2}$ fields.
Including the covariant derivative then adds torsion and anticommutator
couplings.
\begin{eqnarray*}
S_{k=0} & = & \intop\bar{\psi}\left(\frac{1}{2}\boldsymbol{\Gamma}^{1}\land i\overleftrightarrow{\mathbf{d}}-m\boldsymbol{\Gamma}^{0}\right)\psi-\frac{i}{4}\bar{\psi}\left\{ \boldsymbol{\Gamma}^{1},\sigma^{ef}\right\} \psi\land\boldsymbol{\omega}_{ef}\\
S_{k=1} & = & \intop\bar{\boldsymbol{\psi}}\land\left(\boldsymbol{\Gamma}^{3}\land i\mathbf{d}\boldsymbol{\psi}-m\boldsymbol{\Gamma}^{2}\land\boldsymbol{\psi}\right)\\
 &  & -\frac{i}{2}\intop\left(\bar{\boldsymbol{\psi}}\land\boldsymbol{\Gamma}^{3}\land\mathbf{T}^{a}\psi_{a}+\left(-1\right)^{n+1}\mathbf{T}^{a}\bar{\psi}_{a}\land\boldsymbol{\Gamma}^{3}\wedge\boldsymbol{\psi}\right)\\
 &  & -\frac{i}{4}\intop\bar{\boldsymbol{\psi}}\land\left\{ \boldsymbol{\Gamma}^{3},\sigma^{cd}\right\} \land\boldsymbol{\omega}_{cd}\land\boldsymbol{\psi}
\end{eqnarray*}

\subsubsection{General case definitions}

The covariant derivative is similar to that for the Rarita-Schwinger
field (\ref{RS covariant derivative}), but for spin $\frac{2k+1}{2}$
there are is a factor $k$ times the torsion term. Expanding
\[
\boldsymbol{\mathcal{D}}\boldsymbol{\psi}=\mathbf{e}^{a}\wedge\mathbf{e}^{b_{1}}\wedge\ldots\wedge\mathbf{e}^{b_{k}}\mathcal{D}_{a}\psi_{b_{1}\ldots b_{k}}
\]
the expansion is clearest in coordinates,
\begin{eqnarray*}
\boldsymbol{\mathcal{D}}\boldsymbol{\psi} & = & D_{\mu}\psi_{\alpha\ldots\beta}\mathbf{d}x^{\mu}\wedge\mathbf{d}x^{\alpha}\wedge\ldots\wedge\mathbf{d}x^{\beta}\\
 & = & \partial_{\mu}\psi_{\alpha\ldots\beta}\mathbf{d}x^{\mu}\wedge\mathbf{d}x^{\alpha}\wedge\ldots\wedge\mathbf{d}x^{\beta}-\psi_{\rho\ldots\beta}\Gamma_{\;\;\;\alpha\mu}^{\rho}\mathbf{d}x^{\mu}\wedge\mathbf{d}x^{\alpha}\wedge\ldots\wedge\mathbf{d}x^{\beta}\\
 &  & -\ldots-\psi_{\alpha\ldots\rho}\Gamma_{\;\;\;\beta\mu}^{\rho}\mathbf{d}x^{\mu}\wedge\mathbf{d}x^{\alpha}\wedge\ldots\wedge\mathbf{d}x^{\beta}-\frac{1}{2}\omega_{ab\mu}\sigma^{ab}\psi_{\alpha\ldots\beta}\mathbf{d}x^{\mu}\wedge\mathbf{d}x^{\alpha}\wedge\ldots\wedge\mathbf{d}x^{\beta}
\end{eqnarray*}
Antisymmetrizing each $\Gamma_{\;\;\;\alpha\mu}^{\rho}$ gives a torsion
$\psi_{\alpha\ldots\rho}\Gamma_{\;\;\;\beta\mu}^{\rho}\mathbf{d}x^{\mu}\wedge\mathbf{d}x^{\alpha}\wedge\ldots\wedge\mathbf{d}x^{\sigma}\wedge\mathbf{d}x^{\beta}=\mathbf{T}^{\rho}\wedge\boldsymbol{\psi}_{\rho}$
where we define $\boldsymbol{\psi}_{\rho}\equiv\psi_{\rho\alpha\ldots\sigma}\wedge\mathbf{d}x^{\alpha}\wedge\ldots\wedge\mathbf{d}x^{\sigma}$.
We get the same expression for each vector index so rearrangement
gives

\begin{eqnarray}
\boldsymbol{\mathcal{D}}\boldsymbol{\psi} & = & \mathbf{d}\boldsymbol{\psi}-k\mathbf{T}^{\rho}\wedge\boldsymbol{\psi}_{\rho}-\frac{1}{2}\boldsymbol{\omega}_{ab}\wedge\sigma^{ab}\boldsymbol{\psi}\label{Covariant derivative spinor p-form}
\end{eqnarray}
The same result follows in an orthogonal basis, but it is easiest
to see using coordinates.

For the generalized $\Gamma s$ it is useful to normalize to avoid
overall signs. Setting $h\boldsymbol{\Gamma}^{k}=\left(h\boldsymbol{\Gamma}^{k}\right)^{\dagger}$
introduces a factor of $\left(-1\right)^{k}$, but including the fields
the adjoint of the combination $\bar{\boldsymbol{\psi}}\land\boldsymbol{\Gamma}^{2k+1}\land i\mathbf{d}\boldsymbol{\psi}$
introduces an additional factor of $\left(-1\right)^{k}$ . We therefore
require no phase factor and can conveniently define
\begin{eqnarray*}
\boldsymbol{\Gamma}^{m} & \equiv & \frac{1}{m!}\,^{*}\left[\left(\wedge\boldsymbol{\gamma}\right)^{m}\right]
\end{eqnarray*}
for all integers $m$.

\subsubsection{Spin $\frac{2k+1}{2}$ fields}

To start, we take the flat space $Spin\left(\frac{2k+1}{2}\right)$
action to be
\begin{equation}
S_{k}^{0}=\intop\bar{\boldsymbol{\psi}}\land\left(\boldsymbol{\Gamma}^{2k+1}\land i\mathbf{d}\boldsymbol{\psi}-m\boldsymbol{\Gamma}^{2k}\land\boldsymbol{\psi}\right)\label{Real,  flat space action}
\end{equation}
after taking the conjugate and expanding the forms explicitly to check
that $S_{k}^{0}$ is real. Notice that $\bar{\boldsymbol{\psi}}\land\mathbf{d}\boldsymbol{\psi}$
is a $\left(2k+1\right)$-form and therefore $S_{k}^{0}$ exists only
for $n\geq2k+1$. This makes Rarita-Schwinger the maximal case in
4-dimensional spacetime. Then, replacing $\mathbf{d}\Rightarrow\boldsymbol{\mathcal{D}}$
using Eq.(\ref{Covariant derivative spinor p-form}) and symmetrizing,
the gravitationally coupled $Spin\left(\frac{2k+1}{2}\right)$ action
is
\begin{eqnarray*}
S_{k} & = & \frac{1}{2}\left(\tilde{S}_{k}+\tilde{S}_{k}^{*}\right)
\end{eqnarray*}

As with the Rarita-Schwinger case, we find the real part of the torsion
and $\sigma^{ab}$ parts. For the torsion terms
\begin{eqnarray*}
S_{k}\left(T\right) & = & \frac{1}{2}\intop\bar{\boldsymbol{\psi}}\land\left(\boldsymbol{\Gamma}^{2k+1}\land\left(-ik\mathbf{T}^{a}\wedge\boldsymbol{\psi}_{a}\right)\right)+\frac{1}{2}\intop\left[\bar{\boldsymbol{\psi}}\land\boldsymbol{\Gamma}^{2k+1}\land\left(-ik\mathbf{T}^{a}\wedge\boldsymbol{\psi}_{a}\right)\right]^{\dagger}\\
 & = & -\frac{ik}{2}\intop\left(\bar{\boldsymbol{\psi}}\land\boldsymbol{\Gamma}^{2k+1}\land\mathbf{T}^{a}\wedge\boldsymbol{\psi}_{a}+\left(-1\right)^{n+k}\mathbf{T}^{a}\wedge\bar{\boldsymbol{\psi}}_{a}\land\boldsymbol{\Gamma}^{2k+1}\wedge\boldsymbol{\psi}\right)
\end{eqnarray*}
while the $\sigma^{ab}$ terms still give an anticommutator
\begin{eqnarray*}
S_{k}\left(\sigma\right) & = & \frac{1}{2}\intop\bar{\boldsymbol{\psi}}\land\left\{ \boldsymbol{\Gamma}^{2k+1},\sigma^{cd}\right\} \wedge\left(-\frac{i}{2}\boldsymbol{\omega}_{cd}\land\boldsymbol{\psi}\right)
\end{eqnarray*}
Therefore, the action for gravitationally coupled $Spin\left(\frac{2k+1}{2}\right)$
fields is
\begin{eqnarray}
S_{k} & = & \intop\bar{\boldsymbol{\psi}}\land\left(\boldsymbol{\Gamma}^{2k+1}\land i\mathbf{d}\boldsymbol{\psi}-m\boldsymbol{\Gamma}^{2k}\land\boldsymbol{\psi}\right)\nonumber \\
 &  & -\frac{ik}{2}\intop\left(\bar{\boldsymbol{\psi}}\land\boldsymbol{\Gamma}^{2k+1}\land\mathbf{T}^{a}\wedge\boldsymbol{\psi}_{a}+\left(-1\right)^{n+k}\mathbf{T}^{a}\wedge\bar{\boldsymbol{\psi}}_{a}\land\boldsymbol{\Gamma}^{2k+1}\wedge\boldsymbol{\psi}\right)\nonumber \\
 &  & +\frac{1}{2}\intop\bar{\boldsymbol{\psi}}\land\left\{ \boldsymbol{\Gamma}^{2k+1},\sigma^{cd}\right\} \land\left(-\frac{i}{2}\boldsymbol{\omega}_{cd}\boldsymbol{\psi}\right)\label{Action thoroughly checked}
\end{eqnarray}
The spin tensor always contains the anticommutator, which always brings
in couplings involving $\Gamma^{2k-1}$ and $\Gamma^{2k+3}$ only
(see the Appendix). The Dirac field has $k=0$, so only the $\Gamma^{3}$
term is possible, while for Rarita-Schwinger fields with $k=1$ we
see both $\Gamma^{1}$ and $\Gamma^{5}$.

There are also direct torsion couplings of the form
\[
k\bar{\boldsymbol{\psi}}\land\boldsymbol{\Gamma}^{2k+1}\land\mathbf{T}^{a}\wedge\boldsymbol{\psi}_{a}+c.c.
\]
so the $Spin\left(\frac{2k+1}{2}\right)$ field may emit and absorb
torsion. This is absent from Dirac interactions because there is no
vector index on $\psi$, but does show up in the Rarita-Schwinger
spin tensor. If the action includes a dynamical torsion term this
constitutes a new interaction unless there is a consistent interpretation
of torsion in terms of known interactions.

The spin tensor is given by a simple variation, followed by reducing
the basis forms to a volume form. The result is
\begin{eqnarray*}
\sigma^{cab} & = & \frac{ik}{2}\left(-1\right)^{kn-k-n+1}\left(\eta^{be}\delta_{f}^{d}-\left(-1\right)^{k}\eta^{bd}\delta_{f}^{e}\right)\bar{\psi}_{df_{1}\ldots f_{k-1}}\Gamma^{[acff_{1}\ldots f_{k-1}g_{1}\ldots g_{k-1}]}\psi_{eg_{1}\ldots g_{k-1}}\\
 &  & +\frac{i}{4}\left(-1\right)^{kn-k-n+1}\bar{\psi}_{a_{1}\ldots a_{k}}\left\{ \Gamma^{[a_{1}\ldots a_{k}b_{1}\ldots b_{k}c]},\sigma^{ab}\right\} \psi_{b_{1}\ldots b_{k}}\delta_{c_{1}\ldots c_{2k+1}}^{a_{1}\ldots a_{k}b_{1}\ldots b_{k}c}
\end{eqnarray*}
The anticommutator is a linear combination of $\Gamma^{2k-1},\Gamma^{2k+3}$
(See Appendix \ref{sec:Appendix:-Anticommutators}) so together with
the torsion contribution we have the original and both adjacent couplings
$\Gamma^{2k-1},\boldsymbol{\Gamma}^{2k+1},\Gamma^{2k+3}$. It is extremely
likely that, like the Rarita-Schwinger field, higher spin fermions
drive all invariant parts of the torsion.

\subsection{Low dimensions}

We note that the $Spin\left(\frac{2k+1}{2}\right)$ action of Eq.(\ref{Action thoroughly checked})
may be present as long as $n\geq2k+1$. In particular we see Dirac
fields in 2-dimensions, while both Dirac and Rarita-Schwinger fields
may be present in 3-dimensions. We look briefly at these two low dimensional
cases.

\subsubsection{Two dimensions}

In 2-dimensions the anticommutator contributions to the Dirac and
Rarita-Schwinger actions drop out. The actions reduce to
\begin{eqnarray*}
S_{D}^{\left(2\right)} & = & \intop\left(i\bar{\psi}\gamma^{a}\partial_{a}\psi-m\bar{\psi}\psi\right)\boldsymbol{\Phi}\\
S_{RS}^{\left(2\right)} & = & -m\intop\bar{\boldsymbol{\psi}}\land\boldsymbol{\psi}
\end{eqnarray*}
respectively. Neither of these depends on the spin connection, so
there is no source for torsion from ordinary spin-$\frac{2k+1}{2}$
fields.

Ordinary, non-Yang-Mills, bosonic fields can drive 2-dimensional torsion.
For example, setting $\sigma_{c}=\sigma_{c}^{\;\;\;12}$ and $\mathscr{T}_{c}=\mathscr{T}_{c}^{\;\;\;12}$
we might have a vector source of the form
\begin{eqnarray*}
\mathscr{T}_{c}=\sigma_{c} & = & \frac{1}{8}\left(\Theta^{1}D_{c}\Theta^{2}-\varepsilon_{ab}\Theta^{2}D_{c}\Theta^{1}\right)
\end{eqnarray*}
but as we have seen, Yang-Mills sources do not source torsion. For
normal physical fields we therefore expect the usual conformal solutions.
If we do have a non-standard source for torsion, the action is
\begin{eqnarray*}
S & = & \int\left(\mathbf{R}+\mathbf{D}T_{a}\wedge\mathbf{e}^{a}\right)+S_{source}
\end{eqnarray*}
where $T_{a}=T_{a12}$.

\subsubsection{Three dimensions}

In 3-dimensional spacetime general relativity gives vanishing Weyl
curvature and is therefore conformally flat \cite{DeserJackiw'tHooft}
and exactly soluble \cite{Witten}. When a cosmological constant is
included solutions become more general \cite{DeserJackiw} and if
negative permit the BTZ black hole \cite{BTZ}. Coupled to Maxwell
theory still other solutions become possible \cite{Clement}. Alternatively,
torsion has been included in models with topological gravity \cite{MielkeBaekler}.
There has been considerable study of all of these models. 

In Poincarè gauge gravity torsion modifications of the curvature tensor
extend it beyond its one conformal degree of freedom, and will also
permit nontrivial solutions. Dirac fields add one additional degree
of freedom, but we find that Rarita-Schwinger fields are also possible
in 3-dimensions and yield all 9 degrees of freedom of fully general
torsion. We briefly explore the Dirac and Rarita-Schwinger contributions.

With Dirac matrices $\gamma^{0}=\sigma^{3},\gamma^{1}=i\sigma^{1},\gamma^{2}=i\sigma^{2}$
satisfying $\left\{ \gamma^{a},\gamma^{b}\right\} =-2\eta^{ab}$,
so that $\gamma^{[a}\gamma^{b}\gamma^{c]}=+i\varepsilon^{abc}$ the
action becomes
\begin{eqnarray*}
S_{D}^{\left(3\right)} & = & \alpha\intop\left(i\bar{\psi}\gamma^{c}e_{c}^{\;\;\;m}\partial_{m}\psi-m\bar{\psi}\psi\right)\boldsymbol{\Phi}-\alpha\intop i\bar{\psi}\omega_{abc}\gamma^{[c}\gamma^{a}\gamma^{b]}\psi\boldsymbol{\Phi}
\end{eqnarray*}
Varying the spin connection, the torsion is determined by
\begin{eqnarray*}
\frac{\kappa}{2}\mathscr{T}^{cab} & = & \alpha i\bar{\psi}\gamma^{[c}\gamma^{a}\gamma^{b]}\psi\\
 & = & -\alpha\varepsilon^{abc}\bar{\psi}\psi
\end{eqnarray*}
Even if the scalar $\bar{\psi}\psi$ is constant, the $\mathbf{C}_{\;\;\;b}^{c}\wedge\mathbf{C}_{\;\;\;c}^{a}$
term in the curvature contributes an effective cosmological constant,
allowing solutions in 3-dim Poincarè gravity as general as those in
\cite{DeserJackiw,BTZ}, though the energy tensor of the Dirac field
must also be included. Nonconstant values will allow more general
solutions.

In 3-dimensions the spin density for spin-$\frac{3}{2}$ fields reduces
Eq.(\ref{RS spin tensor}) to
\begin{eqnarray*}
\sigma^{cab} & = & -\frac{1}{4}\varepsilon^{cde}\left(\bar{\psi}_{e}\left(\delta_{d}^{a}\psi^{b}-\delta_{d}^{b}\psi^{a}\right)+\left(\bar{\psi}^{b}\delta_{d}^{a}-\bar{\psi}^{a}\delta_{d}^{b}\right)\psi_{e}\right)\\
 &  & +i\left(\bar{\psi}^{a}\gamma^{c}\psi^{b}-\bar{\psi}^{b}\gamma^{c}\psi^{a}\right)+i\left(\eta^{ac}\bar{\psi}^{b}\gamma^{e}\psi_{e}-\eta^{bc}\bar{\psi}^{a}\gamma^{e}\psi_{e}+\eta^{bc}\bar{\psi}_{d}\gamma^{d}\psi^{a}-\eta^{ac}\bar{\psi}_{d}\gamma^{d}\psi^{b}\right)
\end{eqnarray*}
This may be written in terms of two currents:
\begin{eqnarray*}
j^{c\left[ab\right]} & = & \varepsilon^{cde}\bar{\psi}_{e}\delta_{d}^{[a}\psi^{b]}\\
J_{\;\;\;ab}^{c} & = & 2i\bar{\psi}_{[a}\gamma^{c}\psi_{b]}
\end{eqnarray*}
as simply
\begin{eqnarray*}
\sigma^{cab} & = & -Re\left(j^{cab}\right)+J^{cab}-\eta^{ac}J_{e}^{\;\;\;eb}+\eta^{bc}J_{e}^{\;\;\;ea}
\end{eqnarray*}
Given the 12 degrees of freedom of $\psi_{a}$, this appears sufficiently
general to drive all components of the torsion.

\section{Conclusions}

We implemented Poincarè gauging in arbitrary dimension $n$ and signature
$\left(p,q\right)$ using Cartan's methods. The principal fields are
the curvature and torsion 2-forms, given in terms of the solder form
and local Lorentz spin connection. The inclusion of torsion produces
a Riemann-Cartan geometry rather than Riemannian. We displayed the
Bianchi identities and showed that the Riemann-Cartan identities hold
if and only if the Riemannian Bianchi identities hold.

Replicating familiar results, we reproduced general relativity in
Riemannian geometry by setting the torsion to zero and varying only
the metric. The resulting Riemannian geometry is known to be consistent
and metric variation leads to a symmetric energy tensor.

We examined sources for the ECSK theory, that is, the gravity theory
in Riemann-Cartan geometry found by using the Einstein-Hilbert form
of the action with the Einstein-Cartan curvature tensor. The vacuum
theory agrees with general relativity even when both the solder form
and connection are varied independently, but there are frequently
nonvanishing matter sources for both the Einstein tensor and the torsion.

The first issue we dealt with in depth was the choice of independent
variables. The spin connection is the sum of the solder-form-compatible
connection and the contorsion tensor $\boldsymbol{\omega}_{\;\;\;b}^{a}=\boldsymbol{\alpha}_{\;\;\;b}^{a}+\mathbf{C}_{\;\;\;b}^{a}$.
We compared and constrasted the resulting two allowed sets of independent
variables: the solder form and spin connection $\left(\mathbf{e}^{a},\boldsymbol{\omega}_{\;\;\;b}^{a}\right)$
on the one hand and the solder form and the contorsion tensor $\left(\mathbf{e}^{a},\mathbf{C}_{\;\;\;b}^{a}\right)$
on the other. When choosing the latter pair the compatible part of
the spin connection $\boldsymbol{\alpha}_{\;\;\;b}^{a}$ must be treated
through its dependence on the solder form. We demonstrated explicitly
how the two choices of independent variable differ in their relationship
to the Lorentz fibers of the Riemann-Cartan space.

Changing independent variables changes the energy tensor. We showed
that the difference between these two choices leads to the difference
between the (asymmetric) canonical energy tensor and the (symmetric)
Belinfante-Rosenfield energy tensor. When the field equations are
combined both methods yield the same reduced system.

Our main contribution was a more thorough analysis of sources for
torsion. Many, perhaps most, of the research on ECSK theory or its
generalizations to include dynamical torsion have restricted attention
to Dirac fields as sources. This yields a single axial current and
totally antisymmetric torsion. This amounts to only $n$ of the $\frac{1}{2}n^{2}\left(n-1\right)$
degrees of freedom of the torsion.

We took the opposite approach, considering fields of \emph{all} spin.
Only scalar and Yang-Mills fields fail to determine nonvanishing torsion.
In addition to these we looked at symmetric bosonic kinetic forms
and found all to provide sources for torsion. We studied Dirac and
Rarita-Schwinger fields in greater depth. After reproducing the well-known
result for Dirac fields, we developed formalism to describe the spin-$\frac{3}{2}$
Rarita-Schwinger field in arbitrary dimension. Surprisingly, in addition
to dependence on the anticommutator of three gammas with the spin
generator, $\left\{ \gamma^{[a}\gamma^{b}\gamma^{c]},\sigma^{de}\right\} $,
there is a direct coupling to torsion, $\psi_{a}\mathbf{T}^{a}$.
Continuing, we showed that Rarita-Schwinger fields drive all three
independent parts of the torsion: the trace, the totally antisymmetric
part, and the traceless, mixed-symmetry residual. Except in dimensions
$5,7,$ and $9$ the Rarita-Schwinger field has enough degrees of
freedom to produce generic torsion. 

Finally, we looked at higher and lower dimensional cases, generalizing
to spin-$\frac{2k+1}{2}$ sources for all $k$. Like the Rarita-Schwinger
case, these have direct torsion couplings in addition to an anticommutator
$\left\{ \gamma^{[a_{1}}\gamma^{a_{2}}\ldots\gamma^{a_{2k+1}]},\sigma^{de}\right\} $,
and appear to drive all components of the torsion. Specializing to
3-dimensions, the we find the geometric structure substantially enhanced
if Rarita-Schwinger fields are present. \smallskip{}

Acknowledgment: The author wishes to thank Joshua Leiter for numerous
discussions, including the Gibbons-Hawking-York boundary term and
the independent parts of the torsion \cite{Leiter}.

\section*{Appendix: Anticommutators\label{sec:Appendix:-Anticommutators}}

The anticommutator of the generators $\sigma^{de}$ with any odd number
of antisymmetrized $\gamma s$ has the form
\[
\left\{ \gamma^{[a_{1}}\gamma^{a_{2}}\ldots\gamma^{a_{2k+1}]},\sigma^{de}\right\} 
\]
We may simplify this by specifying that the $a_{i}$ are all different
and $d\neq e$. Then we have $\sigma^{de}=2\gamma^{d}\gamma^{e}$
and may rearrange the $a_{i}$ in increasing order $a_{1}\leq a_{i}<a_{j}\leq a_{k}$
with the appropriate sign.

With these conditions there are three cases. First, if neither $d$
nor $e$ equals any of the $a_{i}$ then the full product is antisymmetric.
\[
\left\{ \gamma^{[a_{1}}\ldots\gamma^{a_{2k+1}]},\sigma^{de}\right\} =4\gamma^{[a_{1}}\ldots\gamma^{a_{3}}\gamma^{d}\gamma^{e]}
\]
For the second case, suppose exactly one of $e,d$ equals one of the
$a_{i}$. All other $\gamma s$ anticommute. Without loss of generality
let $a_{i}=d$ with $e$ distinct. Then with $\gamma^{d}\gamma^{d}=-\eta^{dd}$
we have
\begin{eqnarray*}
\left\{ \gamma^{[a_{1}}\ldots\gamma^{a_{i}}\ldots\gamma^{a_{2k+1}]},\sigma^{de}\right\}  & = & 2\gamma^{a_{1}}\ldots\gamma^{a_{i}}\ldots\gamma^{a_{2k+1}}\gamma^{d}\gamma^{e}+2\gamma^{d}\gamma^{e}\gamma^{a_{1}}\ldots\gamma^{a_{i}}\ldots\gamma^{a_{2k+1}}\\
 & = & \left(-1\right)^{2k+1-i}2\gamma^{a_{1}}\ldots\gamma^{a_{i}}\gamma^{d}\ldots\gamma^{a_{k}}\gamma^{e}+\left(-1\right)^{i}2\gamma^{e}\gamma^{a_{1}}\ldots\gamma^{d}\gamma^{a_{i}}\ldots\gamma^{a_{k}}\\
 & = & \left(-1\right)^{i}\left(-2\gamma^{a_{1}}\ldots\gamma^{a_{i}}\gamma^{d}\ldots\gamma^{a_{k}}\gamma^{e}+2\gamma^{e}\gamma^{a_{1}}\ldots\gamma^{d}\gamma^{a_{i}}\ldots\gamma^{a_{k}}\right)\\
 & = & 0
\end{eqnarray*}
and this vanishes for all $k$.This holds for every instance.

The only reduced term that can occur is when both $d$ and $e$ match
some $a_{i},a_{j}$. There are two cases: $d=a_{i},e=a_{j}$ and $d=a_{j},e=a_{i}$
where $i<j$. For the first case 
\begin{eqnarray*}
\left\{ \gamma^{[a_{1}}\ldots\gamma^{a_{2k+1}]},\sigma^{de}\right\}  & = & 2\gamma^{a_{1}}\ldots\gamma^{a_{i}}\ldots\gamma^{a_{j}}\ldots\gamma^{a_{2k+1}}\gamma^{d}\gamma^{e}+2\gamma^{d}\gamma^{e}\gamma^{a_{1}}\ldots\gamma^{a_{i}}\ldots\gamma^{a_{j}}\ldots\gamma^{a_{2k+1}}\\
 & = & 2\gamma^{a_{1}}\ldots\left(-1\right)^{2k+1-i}\gamma^{a_{i}}\gamma^{d}\ldots\left(-1\right)^{2k+1-j}\gamma^{a_{j}}\gamma^{e}\ldots\gamma^{a_{2k+1}}\\
 &  & +2\gamma^{a_{1}}\ldots\left(-1\right)^{i-1}\gamma^{d}\gamma^{a_{i}}\ldots\left(-1\right)^{j-1}\gamma^{e}\gamma^{a_{j}}\ldots\gamma^{a_{2k+1}}\\
 & = & \left(-1\right)^{i+j}4\eta^{a_{i}d}\eta^{a_{j}e}\gamma^{a_{1}}\ldots\underset{\land}{\gamma^{a_{i}}}\ldots\underset{\land}{\gamma^{a_{j}}}\ldots\gamma^{a_{2k+1}}
\end{eqnarray*}
where $\underset{\land}{\gamma^{a_{i}}}$ indicates omission of $\gamma^{a_{i}}$.

For the second case, $d=a_{j},e=a_{i}$, we simply exchange $\sigma^{de}=-\sigma^{ed}$
so we just replace $\eta^{a_{i}d}\eta^{a_{j}e}\rightarrow-\eta^{a_{i}e}\eta^{a_{j}d}$.
Combining both terms
\begin{eqnarray*}
\left\{ \gamma^{[a_{1}}\ldots\gamma^{a_{k}]},\sigma^{de}\right\}  & = & 4\left(\eta^{a_{i}d}\eta^{a_{j}e}-\eta^{a_{j}d}\eta^{a_{i}e}\right)\left(-1\right)^{i+j}\gamma^{[a_{1}}\ldots\underset{\land}{\gamma^{a_{i}}}\ldots\underset{\land}{\gamma^{a_{j}}}\ldots\gamma^{a_{k}]}
\end{eqnarray*}
and we have a term like this for each $a_{1}\leq a_{i}<a_{j}\leq a_{k}$.

Therefore, for any set of fixed $a_{1}<a_{2}<\ldots<a_{k}$ and $d<e$,
the general result is
\begin{eqnarray*}
\left\{ \gamma^{[a_{1}}\ldots\gamma^{a_{k}]},\sigma^{de}\right\}  & = & 4\gamma^{[a_{1}}\ldots\gamma^{a_{k}}\gamma^{d}\gamma^{e]}\\
 &  & +\,4\sum_{a_{1}\leq a_{i}<a_{j}\leq a_{k}}\left(\eta^{a_{i}d}\eta^{a_{j}e}-\eta^{a_{j}d}\eta^{a_{i}e}\right)\left(-1\right)^{i+j}\gamma^{[a_{1}}\ldots\underset{\land}{\gamma^{a_{i}}}\ldots\underset{\land}{\gamma^{a_{j}}}\ldots\gamma^{a_{k}]}
\end{eqnarray*}
The essential feature here is that the anticommutator coupling between
$\sigma^{de}$ and $\Gamma^{2k+1}$ always leads to a linear combination
of $\Gamma^{2k+3}$ and $\Gamma^{2k-1}$ and only these.

\begin{thebibliography}{10}
\bibitem{O'Raifeartaigh}L. O\textquoteright Raifeartaigh, The Dawning
of Gauge Theory, Princeton Series in Physics, Princeton U. Press,
Princeton (1997)

\bibitem{Utiyama}Utiyama, Ryoyo, Phys. Rev. 101 (1956) 1597.

\bibitem{Sciama}Sciama, D. W. (1964-01-01). \emph{The Physical Structure
of General Relativity}, Reviews of Modern Physics. 36 (1): 463--469.
Bibcode:1964RvMP...36..463S. doi:10.1103/revmodphys.36.463. ISSN 0034-6861.

\bibitem{Kibble 1961}Kibble, T. W. B. (1961). \textquotedbl Lorentz
Invariance and the Gravitational Field\textquotedbl . Journal of
Mathematical Physics. 2 (2): 212--221. Bibcode:1961JMP.....2..212K.
doi:10.1063/1.1703702. ISSN 0022-2488. S2CID 54806287.

\bibitem{KobayashiNomizu I}Kobayashi, S.; Nomizu, K. (2009) {[}1963{]}.
Foundations of Differential Geometry. Wiley Classics Library. Vol.
1. Wiley. ISBN 978-0-471-15733-5. Zbl 0119.37502.

\bibitem{KobayashiNomizu II}Kobayashi, S.; Nomizu, K. (2009) {[}1969{]}.
Foundations of Differential Geometry. Wiley Classics Library. Vol.
2. Wiley. ISBN 978-0-471-15732-8. Zbl 0175.48504.

\bibitem{Ne'emanRegge}Ne\textquoteright eman, Y. and Regge, T., \emph{Gravity
And Supergravity As Gauge Theories On A Group Manifold}, Phys. Lett.
B 74 (1978) 54.

\bibitem{Ne'emanRegge2}Ne\textquoteright eman, Y. and Regge, T.,
\emph{Gauge Theory Of Gravity And Supergravity On A Group Manifold},
Riv. Nuovo Cim. 1N5 (1978), 1.

\bibitem{IvanovNiederle}Ivanov, E.A. and Niederle, J., \emph{Gauge
formulation of gravitation theories. I}. The Poincaré, de Sitter,
and conformal cases, Phys. Rev. D 25 4 (1982) 976.

\bibitem{IvanovNiederleII}Ivanov, E.A. and Niederle, J., \emph{Gauge
formulation of gravitation theories. II. The special conformal case},
Phys. Rev. D 25 4 (1982) 988.

\bibitem{Wheeler2014}Wheeler, James Thomas, \emph{Gauge theories
of general relativity}, 24th Midwest Relativity Meeting (2014) Available
at: http://works.bepress.com/james\_wheeler/102/

\bibitem{Palatini}Palatini, A., Deduzione invariantiva delle equazioni
gravitazionali dal principio di Hamilton. Rend. Circ. Matem. Palermo
43, 203--212 (1919). https://doi.org/10.1007/BF03014670

\bibitem{Cartan 1922}Cartan, Élie, \emph{Sur une généralisation de
la notion de courbure de Riemann et les espaces à torsion}, Comptes
rendus de l'Académie des Sciences de Paris (in French) 174: (1922)
593--595.

\bibitem{Cartan 1923}Cartan, Elie, \emph{Sur les variétés à connexion
affine et la théorie de la relativité généralisée (première partie)},
Annales Scientifiques de l'École Normale Supérieure (in French), 40:
(1923) 325--412. doi:10.24033/asens.751. ISSN 0012-9593.

\bibitem{Cartan 1924}Cartan, Elie, \emph{Sur les variétés à connexion
affine, et la théorie de la relativité généralisée (première partie)
(Suite)}, Annales Scientifiques de l'École Normale Supérieure (in
French). 41: (1924). 1--25. doi:10.24033/asens.753. ISSN 0012-9593.

\bibitem{Cartan 1925}Cartan, Elie, \emph{Sur les variétés à connexion
affine, et la théorie de la relativité généralisée (deuxième partie)},
Annales Scientifiques de l'École Normale Supérieure (in French), 42:
(1925) 17--88. doi:10.24033/asens.761. ISSN 0012-9593.

\bibitem{Einstein 1928}Einstein, Albert (1928). \emph{Riemann-Geometrie
mit Aufrechterhaltung des Begriffes des Fernparallelismus}, Preussische
Akademie der Wissenschaften, Phys.-math. Klasse, Sitzungsberichte,
1928: 217--221.

\bibitem{HehlNester}Hehl, Friedrich W.; von der Heyde, Paul; Kerlick,
G. David; Nester, James M. (1976-07-01). \emph{General relativity
with spin and torsion: Foundations and prospects}. Reviews of Modern
Physics. 48 (3): 393--416. doi:10.1103/revmodphys.48.393

\bibitem{Neville 1980}Donald E. Neville, \emph{Gravity theories with
propagating torsion}, Phys. Rev. D 21, 867 -- Published 15 February
1980

\bibitem{CarrollField}Carroll, Sean M. and George B. Field, \emph{Consequences
of Propagating Torsion in Connection-Dynamic Thories of Gravity},
Phys.Rev. D50 (1994) 3867-3873, arXiv:gr-qc/9403058v2. DOI: 10.1103/PhysRevD.50.3867

\bibitem{SezginvanNieuwenhuizen}E. Sezgin and P. van Nieuwenhuizen,
\emph{New ghost-free gravity Lagrangians with propagating torsion},
Phys. Rev. D 21, 3269 -- Published 15 June 1980

\bibitem{Saa}Saa, A. \emph{Propagating Torsion from First Principles}.
General Relativity and Gravitation 29, 205--220 (1997). https://doi.org/10.1023/A:1010240011895

\bibitem{BelyaevShapiro}A.S. Belyaev, Ilya L. Shapiro, \emph{The
action for the (propagating) torsion and the limits on the torsion
parameters from present experimental data}, Physics Letters B Volume
425, Issues 3--4, 23 April 1998, Pages 246-254

\bibitem{Shapiro}I.L. Shapiro, \emph{Physical Aspects of the Space-Time
Torsion, }Phys.Rept.357:113,2002, arXiv:hep-th/0103093, https://doi.org/10.48550/arXiv.hep-th/0103093 

\bibitem{HayashiNakano}K. Hayashi and T. Nakano, Prog. Theor. Phys.
38 (1967), 491.

\bibitem{Wheeler2023}Wheeler, James T., \emph{Internal symmetry in
Poincare gauge theory}, arXiv:2304.14586 {[}hep-th{]}, https://doi.org/10.48550/arXiv.2304.14586

\bibitem{York1972}York, J. W. (1972). \textquotedbl\emph{Role of
conformal three-geometry in the dynamics of gravitation}\textquotedbl .
Physical Review Letters. 28 (16): 1082. Bibcode:1972PhRvL..28.1082Y.
doi:10.1103/PhysRevLett.28.1082.

\bibitem{GibbonsHawking}Gibbons, G. W.; Hawking, S. W. (1977). \textquotedbl\emph{Action
integrals and partition functions in quantum gravity}\textquotedbl .
Physical Review D. 15 (10): 2752. Bibcode:1977PhRvD..15.2752G. doi:10.1103/PhysRevD.15.2752.

\bibitem{HawkingHorowitz}Hawking, S W; Horowitz, Gary T (1996-06-01).
\textquotedbl\emph{The gravitational Hamiltonian, action, entropy
and surface terms}\textquotedbl . Classical and Quantum Gravity.
13 (6): 1487--1498. arXiv:gr-qc/9501014. Bibcode:1996CQGra..13.1487H.
doi:10.1088/0264-9381/13/6/017. ISSN 0264-9381. S2CID 12720010.

\bibitem{BrownYork}Brown, J. David; York, James W. (1993-02-15).
\textquotedbl\emph{Microcanonical functional integral for the gravitational
field}\textquotedbl . Physical Review D. American Physical Society
(APS). 47 (4): 1420--1431. arXiv:gr-qc/9209014. Bibcode:1993PhRvD..47.1420B.
doi:10.1103/physrevd.47.1420. ISSN 0556-2821. PMID 10015718. S2CID
25039417.

\bibitem{Belinfante}F. J. Belinfante (1940). \textquotedbl\emph{On
the current and the density of the electric charge, the energy, the
linear momentum and the angular momentum of arbitrary fields}\textquotedbl .
Physica. 7 (5): 449. Bibcode:1940Phy.....7..449B. CiteSeerX 10.1.1.205.8093.
doi:10.1016/S0031-8914(40)90091-X.

\bibitem{Rosenfeld}L. Rosenfeld (1940). \textquotedbl\emph{Sur
le tenseur D'Impulsion--Energie}\textquotedbl . Acad. Roy. Belg.
Memoirs de Classes de Science. 18 (fasc. 6).

\bibitem{HenneauxTeitelboim}Henneaux, M., Teitelboim, C. p-Form electrodynamics.
Found Phys 16, 593--617 (1986). https://doi.org/10.1007/BF01889624

\bibitem{Proca}Proca, A., \emph{Sur la théorie ondulatoire des electrons
positifs et négatifs,} J. Phys. Radium 7, 347--353 (1936); \emph{Sur
la théorie du positon.} C. R. Acad. Sci. Paris 202, 1366 (1936); \emph{Sur
les equations fondamentales des particules elémentaires}, C. R. Acad.
Sci. Paris 202, 1490 (1936); \emph{Théorie non relativiste des particulés
a spin entier}, J. Phys. Radium 9, 61 (1938).

\bibitem{Datta}Datta, \emph{Spinor fields in general relativity,
II: Generalized field equations and application to the Dirac field},
B. K. Il Nuovo Cimento B (1971-1996) volume 6, pages 16--28 (1971)

\bibitem{HehlDatta}F. W. Hehl and B. K. Datta, \emph{Nonlinear Spinor
Equation and Asymmetric Connection in General Relativity,} J. Math.
Phys. 12, 1334 (1971); https://doi.org/10.1063/1.1665738 

\bibitem{HayashiBreegman}K Hayashi, A Bregman, \emph{Poincaré gauge
invariance and the dynamical role of spin in gravitational theory},
Annals of Physics Volume 75, Issue 2, (February 1973), Pages 562-600,
Received 30 November 1971.

\bibitem{Hehl}F. W. Hehl, \emph{Spin and torsion in general relativity:
I. Foundations}, General Relativity and Gravitation volume 4, pages
333--349 (1973)

\bibitem{HehlvonderHeydeKerlick}Friedrich W. Hehl, Paul von der Heyde,
and G. David Kerlick, \emph{General relativity with spin and torsion
and its deviations from Einstein's theory}, Phys. Rev. D 10, (15 August
1974) 1066. 

\bibitem{HehlNitschvonderHeyde}F. W. Hehl, J. Nitsch and P. von der
Heyde, in General Relativity and Gravitation, ed. A. Held (Plenum
Press, New York, 1980).

\bibitem{RaritaSchwinger}William Rarita and Julian Schwinger. \textquotedblleft \emph{On
a Theory of Particles with Half-Integral Spin}\textquotedblright .
Phys. Rev. 60 (1941), 61--61.

\bibitem{Buchdahl}Buchdahl, H.A. \emph{On the compatibility of relativistic
wave equations for particles of higher spin in the presence of a gravitational
field}. Nuovo Cim 10, 96--103 (1958). https://doi.org/10.1007/BF02859608

\bibitem{Freedman vanN Rarita Sch}D.Z. Freedman, P. van Nieuwenhuizen,
Phys. Rev. D14 (1976) 912

\bibitem{Freedman1976}D.Z. Freedman, P. van Nieuwenhuizen, S. Ferrara,
\emph{Progress toward a theory of supergravity}, Phys. Rev. D13 (1976)
3214

\bibitem{MacDowellMansouri}MacDowell, S. W.; Mansouri, F. (1977),
\emph{Unified geometric theory of gravity and supergravity}. Phys.
Rev. Lett. 38 (14): 739--742. doi:10.1103/PhysRevLett.38.739. 

\bibitem{CremmerJuliaScherk}Cremmer, E.; Julia, B.; Scherk, J. (1978).
\emph{Supergravity in theory in 11 dimensions}. Physics Letters B.
Elsevier BV. 76 (4): 409--412. doi:10.1016/0370-2693(78)90894-8.
ISSN 0370-2693.

\bibitem{Passiaas11D}Achilleas Passias, \emph{Aspects of Supergravity
in Eleven Dimensions}, MS Thesis, Imperial College London, September
2010.

\bibitem{Leiter}Joshua Leiter, Utah State University PhD dissertation
(2023).

\bibitem{DeserJackiw'tHooft}Stanley Deser, R. Jackiw, Gerard 't Hooft,
\emph{Three-Dimensional Einstein Gravity: Dynamics of Flat Space},
Annals Phys. 152 (1984) 220 DOI: 10.1016/0003-4916(84)90085-X

\bibitem{DeserJackiw}Stanley Deser, R. Jackiw, \emph{Three-Dimensional
Cosmological Gravity: Dynamics of Constant Curvature}, Annals Phys.
153 (1984), 405-416 DOI: 10.1016/0003-4916(84)90025-3

\bibitem{Witten}Edward Witten, \emph{(2+1)-Dimensional Gravity as
an Exactly Soluble System,} Nucl.Phys.B 311 (1988) 46, DOI: 10.1016/0550-3213(88)90143-5

\bibitem{MielkeBaekler}Eckehard W. Mielke, Peter Baekler, \emph{Topological
gauge model of gravity with torsion}, Phys.Lett.A 156 (1991) 399-403
DOI: 10.1016/0375-9601(91)90715-K

\bibitem{Clement}G Clement, \emph{Classical solutions in three-dimensional
Einstein-Maxwell cosmological gravity}, Classical and Quantum Gravity,
Vol. 10, N. 5 (1993) DOI 10.1088/0264-9381/10/5/002

\bibitem{BTZ}Máximo Bañados, Claudio Teitelboim, and Jorge Zanelli,
\emph{Black hole in three-dimensional spacetime,} Phys. Rev. Lett.
69, (28 Sept. 1992) 1849, DOI:https://doi.org/10.1103/PhysRevLett.69.1849
\end{thebibliography}
\end{document}